\documentclass[twocolumn]{aastex63}
\usepackage[encapsulated]{CJK}
\usepackage{mathrsfs}
\usepackage{subfigure}
\usepackage{epstopdf}
\usepackage{amsmath}
\usepackage{longtable}
\usepackage{booktabs}
\usepackage{hyperref} 
\usepackage{overpic}
\usepackage{lineno}

\shorttitle{Rotation in Praesepe}
\shortauthors{Hao et al.}

\definecolor{malachite}{rgb}{0.34, 0.7, 0.22}

\begin{document}

   \title{On the nature of rotation in the Praesepe cluster}
   
   \correspondingauthor{Ye Xu, Ligang Hou}
   \email{xuye@pmo.ac.cn, lghou@nao.cas.cn}
   
   \author{C. J. Hao}
   \affiliation{Purple Mountain Observatory, Chinese Academy of Sciences, Nanjing 210023, PR China}
   \affiliation{School of Astronomy and Space Science, University of Science and Technology of China, Hefei 230026, PR China}
   \author{Y. Xu}
   \affiliation{Purple Mountain Observatory, Chinese Academy of Sciences, Nanjing 210023, PR China}
   \affiliation{School of Astronomy and Space Science, University of Science and Technology of China, Hefei 230026, PR China}
   \author{S. B. Bian}
   \affiliation{Purple Mountain Observatory, Chinese Academy of Sciences, Nanjing 210023, PR China}
   \affiliation{School of Astronomy and Space Science, University of Science and Technology of China, Hefei 230026, PR China}
   \author{L. G. Hou}
   \affiliation{National Astronomical Observatories, Chinese Academy of Sciences, 20A Datun Road, Chaoyang District, Beijing 100101, PR China}
   \author{Z. H. Lin}
   \affiliation{Purple Mountain Observatory, Chinese Academy of Sciences, Nanjing 210023, PR China}
   \affiliation{School of Astronomy and Space Science, University of Science and Technology of China, Hefei 230026, PR China}
   \author{Y. J. Li}
   \affiliation{Purple Mountain Observatory, Chinese Academy of Sciences, Nanjing 210023, PR China}
   \author{D. J. Liu}
   \affiliation{Purple Mountain Observatory, Chinese Academy of Sciences, Nanjing 210023, PR China}
   \affiliation{School of Astronomy and Space Science, University of Science and Technology of China, Hefei 230026, PR China}

 
\begin{abstract}
Although a large number of Galactic open clusters (OCs) have been
identified, the internal kinematic properties (e.g., rotation) of
almost all the known OCs are still far from clear.
With the high-precision astrometric data of {\it Gaia} EDR3, we 
have developed a methodology to unveil the rotational properties 
of the Praesepe cluster.
Statistics of the three-dimensional residual motions of the member 
stars reveal the presence of Praesepe's rotation and determine 
its spatial rotation axis.
The mean rotation velocity of the Praesepe cluster within its
tidal radius is estimated to be 0.2 $\pm$ 0.05~km~s$^{-1}$, 
and the corresponding rotation axis is tilted in relation to the Galactic 
plane with an angle of $41^\circ\pm12^\circ$.
We also analysed the rms rotational velocity
of the member stars around the rotation axis, and found that the
rotation of the member stars within the tidal radius of Praesepe
probably follows the Newton’s classical theorems.
\end{abstract}

\keywords{open clusters and associations: individual (Praesepe)
--  stars: kinematics and dynamics -- methods: statistical}

%

\section{Introduction}
\label{intro}

The open star clusters in the Galaxy are gravitationally bound
systems, and present a wide range of ages from a few million 
years to several billion years.
An open cluster (OC) typically contains tens to thousands of member stars, 
which formed almost simultaneously in the same molecular
cloud~\citep[e.g.,][]{lada2003}.
OCs serve as excellent astronomical laboratories for studying the
stellar structure and evolution~\citep[e.g.,][]{Barnes2007,Motta2017,
marino2018}, and are good tracers for unveiling the structure 
and evolution of our Galaxy~\citep[][]{castro2021,poggio2021,
hao2021,hou2021}.
Although thousands of OCs have been identified in the Milky
Way~\citep[][]{dias2002,kharchenko2013,
cantat2020,castro2020,hao2022,castro2022},
little is known about their internal kinematics, such as rotation and 
whether they are expanding or contracting.
These kinematic properties are strongly related to the formation and
evolution of OCs.

In the early days, there were some efforts focused on investigating the
effect of cluster rotation on the proper motion vector geometry of its
member stars~\citep[e.g.,][]{wayman1967,hanson1975,gunn1988,
perryman1998}.
Later, based on the dataset provided by the $Hipparcos$ and WEBDA,
\citet{vereshchagin2013a} and \citet{vereshchagin2013b} attempted to
explore the potential rotation in the Hyades and Praesepe clusters.
They found a correlation between the tangential velocities and the
parallaxes of cluster members, which implies the possible rotation in 
the two OCs. 
Recently, by taking advantage of the released dataset of 
{\it Gaia}~\citep{prusti2016}, some works studied the relationship 
between the tangential or radial
velocities (RVs) and the angular radii for the member stars of some
OCs~\citep[e.g.,][]{kamann2019,Healy2021}. The signals of
rotation were found in the Praesepe and NGC 6791 clusters, but not 
for the Pleiades and NGC 6819 clusters.
In brief, the kinematic properties have only been simply explored for 
several OCs, which remain as major omission in the studies of Galactic 
OCs.

In this study, we focus on the Praesepe cluster, also known as the
Beehive Cluster, located at a distance of 170$--$190~pc from the
Sun~\citep[e.g.,][]{pinsonneault1998,dias2002,kharchenko2013,
cantat2018,babusiaux2018,gao2019}.
It is an intermediate-age cluster with an estimated age of 590–660
Myr~\citep[e.g.,][]{mermilliod1981,vandenberg1984,delorme2011,
brandt2015,gossage2018}.
The tidal radius of the Praesepe cluster is estimated to be in the
range of [10, 12]
pc~\citep[e.g.,][]{adams2002,khalaj2013,lodieu2019,roser2019,
gao2019,loktin2020}. The member stars within the tidal radius of the
cluster are generally gravitationally bound.
As a rich OC, Praesepe contains red giants and white dwarfs
representing the late stages of stellar evolution, as well as
main-sequence stars and a significant number of low-mass
stars~\citep[e.g.,][]{dobbie2004,dobbie2006,kraus2007,wang2014,
babusiaux2018}.
\cite{kleinwassink1924,kleinwassink1927}~firstly studied this cluster.
The low-mass members, binarity, luminosity and mass functions of the
Praesepe cluster were then investigated by the subsequent
works~\citep[see][and references within]{lodieu2019}.
Recently, with the {\it Gaia} data, \cite{roser2019}~found a clear indication
for the existence of tidal tails in Praesepe, even stretching to about
165~pc from the cluster centre.

Based on the trigonometric parallaxes, proper motions, and RVs 
in {\it Gaia} data release 2~\citep[DR2,][]{brown2018}, 
\cite{loktin2020}~reported a possible rotation at the periphery 
of Praesepe with a velocity of 0.4~km s$^{ -1}$. 
As the improved data quality of the {\it Gaia} early data release 
3~\citep[EDR3,][]{brown2021}, the detailed kinematic
properties are expected to unveil the nature of Praesepe further.
In contrast to previous studies only using the projection methods, 
this study aims to investigate the global and particularly the 
internal kinematics of Praesepe in three dimensions (3D)
simultaneously, according to the high-precision astrometric 
parameters of {\it Gaia} EDR3.
In this work, we will first extract the member stars of Praesepe from
the {\it Gaia} EDR3, all of which have the six-parameter solutions and
RVs.
Subsequently, we develop a methodology used to study the kinematic
properties of a Galactic OC, including calculating and analysing the
3D residual velocities of the member stars,
determining the rotation axis if the cluster rotates, and deriving the 
rotational velocities of the member stars.
This methodology is applied to the Praesepe cluster to 
characterize its kinematics.

\section{Sample of the member stars}
\label{sample}

\begin{figure*}
\begin{center}
\includegraphics[width=1.00\textwidth]{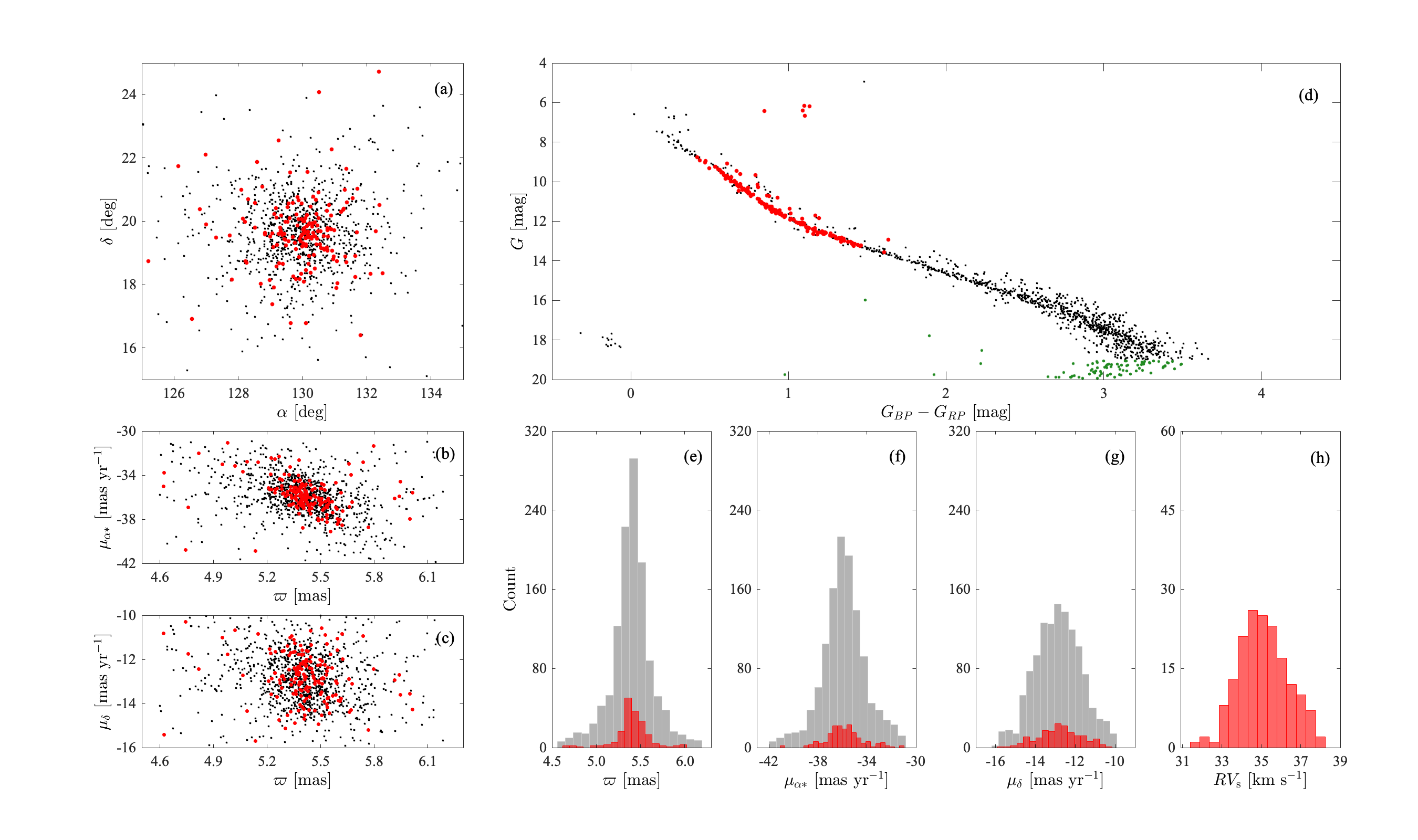}
\caption{Properties of the parameters for the member stars (grey) of
  Praesepe, as well as those for the members with RV
  measurements (red). {\it Panel} (a): distribution on the sky. {\it
    Panels} (b) and (c): diagrams of parallax vs. proper motions. {\it
    Panel} (d): color--magnitude diagram, where the green dots
    represent the excluded stars that are below and far away from the
    main sequence. {\it Panels} (e), (f), (g), and (h): histograms
  of the parallaxes, proper motions, and RVs,
  respectively.}
\label{fig:distribution}
\end{center}
\end{figure*}

The member stars of Praesepe are relatively easy to identify, as the
Praesepe cluster is close to the Sun and presents large proper
motions~\citep[e.g.,][]{pinsonneault1998,dias2002,
kharchenko2013,cantat2018,babusiaux2018}.
To identify the possible members, we adopt a selection method similar
to that of \cite{loktin2020}, but based on the {\it Gaia} EDR3.
First, the stars falling within a radius of $10^\circ$ around the
coordinates of $RA$ (Right ascension) = 08h40m24s and $Dec$ 
(Declination) = $+19^\circ 40^\prime
00^{\prime \prime}$~(J2000) are selected as the initial dataset.
Field stars are then eliminated according to the diagram of
trigonometric parallax ($\varpi$) and apparent magnitude ($G$).
The selection boundary for the trigonometric parallax is set to [4.6,
6.1]~mas, which represents three-times of the mean parallax
uncertainty for the faint stars ($G$ = 20) in {\it Gaia} EDR3.
A similar procedure is applied to the diagrams of proper
motions ($\mu_{\alpha^{*}}$, $\mu_{\delta}$) and apparent 
magnitude, where the selection boundaries are set to be 
[$-$41, $-$31]~mas~yr$^{-1}$ and [$-$16, $-$10]~mas~yr$^{-1}$, 
respectively, by referencing the mean and
three-times standard deviation of $\mu_{\alpha^{*}}$ and
$\mu_{\delta}$ reported by \cite{loktin2020}.
Similar to \citet[][see their Figure 3]{loktin2020}, the stars below
and far away from the main sequence are identified by eye and
excluded according to a color--magnitude diagram, 
as indicated by the green dots in Figure~\ref{fig:distribution}(d),
except for a group of white dwarfs already known to belong to
Praesepe~\citep[e.g.,][]{dobbie2004,dobbie2006,babusiaux2018}. In
addition, we notice that all the excluded objects are faint stars and
do not possess RV measurements in {\it Gaia} EDR3.
After eliminating the stars with uncertainties of trigonometric
parallax or proper motions larger than 10\%, we obtain a catalog
containing 1\,135 member stars. Figure~\ref{fig:distribution}
shows the distributions of the observed parameters (i.e., $RA$,
$Dec$, $\varpi$, $\mu_{\alpha^{*}}$, and $\mu_{\delta}$) of the
member stars, which are approximated by a multidimensional normal
distribution. Objects that deviate from the three-times standard
deviation of the mean values are possible field stars. We found that
only about 5.8\% of the member stars deviate in some of the
five-dimensional space. The influence of contaminating field stars
on the following results is believed to be small.
%

Because the aim of this study is to inspect the 3D kinematic
properties of the Praesepe cluster, only the stars with RV 
measurements are extracted from the catalog.
Here, we adopt only the RVs provided by {\it Gaia} to 
ensure the homogeneity of the sample and avoid the potentially 
significant systematic errors.
Stars with uncertainties of RVs larger than
2~km~s$^{-1}$ are eliminated, in order to ensure an accurate
measurements of RV.
20 outliers are further rejected after checking the member stars
one by one, as they possess RV values that significantly diverge from 
the three-times standard deviation of the reported mean RV of
Praesepe~\citep[$35.1$ $\pm$ $1.6$ km s$^{-1}$,][]{loktin2020}.
In total, a sample of 172
member stars of the Praesepe cluster is selected, as listed in
Table~\ref{table:table_s1} in the Appendix.
If some of the member stars are in binary systems with spurious 
astrometric solutions, they might
affect the kinematics of an OC derived from the astrometric
measurements and RVs of the member stars. For the close binary stars,
the reliability of astrometry and RVs provided by {\it
Gaia} EDR3 has been improved in comparison with {\it Gaia}
DR2~\citep{seabroke2021}. As discussed by \cite{brown2021}, a value of 
the parameter \texttt{ipd\_gof\_harmonic\_amplitude} above 0.1 in
combination with \texttt{ruwe} being larger than 1.4 are indicators of a
source that is non-single and not correctly handled in the astrometric
solution. We do not find such sources in our selected member stars
of Praesepe after inspecting these two parameters.
For the member stars in the sample, the median uncertainties of 
the parallax, proper motions $\mu_{\alpha^{*}}$ and $\mu_{\delta}$ 
are 0.02~mas, 0.02~mas~yr$^{-1}$, and 0.015~mas~yr$^{-1}$, 
respectively.
The median uncertainty for the RV measurements is 0.9~km s$^{ -1}$.

Figure~\ref{fig:distribution} shows the distributions of the
astrometric parameters ($RA$, $Dec$, $\varpi$, $\mu_{\alpha^{*}}$,
$\mu_{\delta}$, and RVs) of the member stars, including the
color-magnitude diagram.
It is shown that the observed parameters of the cluster member stars 
are approximated by a multidimensional normal distribution.
%

\begin{figure}
\begin{center}
\includegraphics[width=0.48\textwidth]{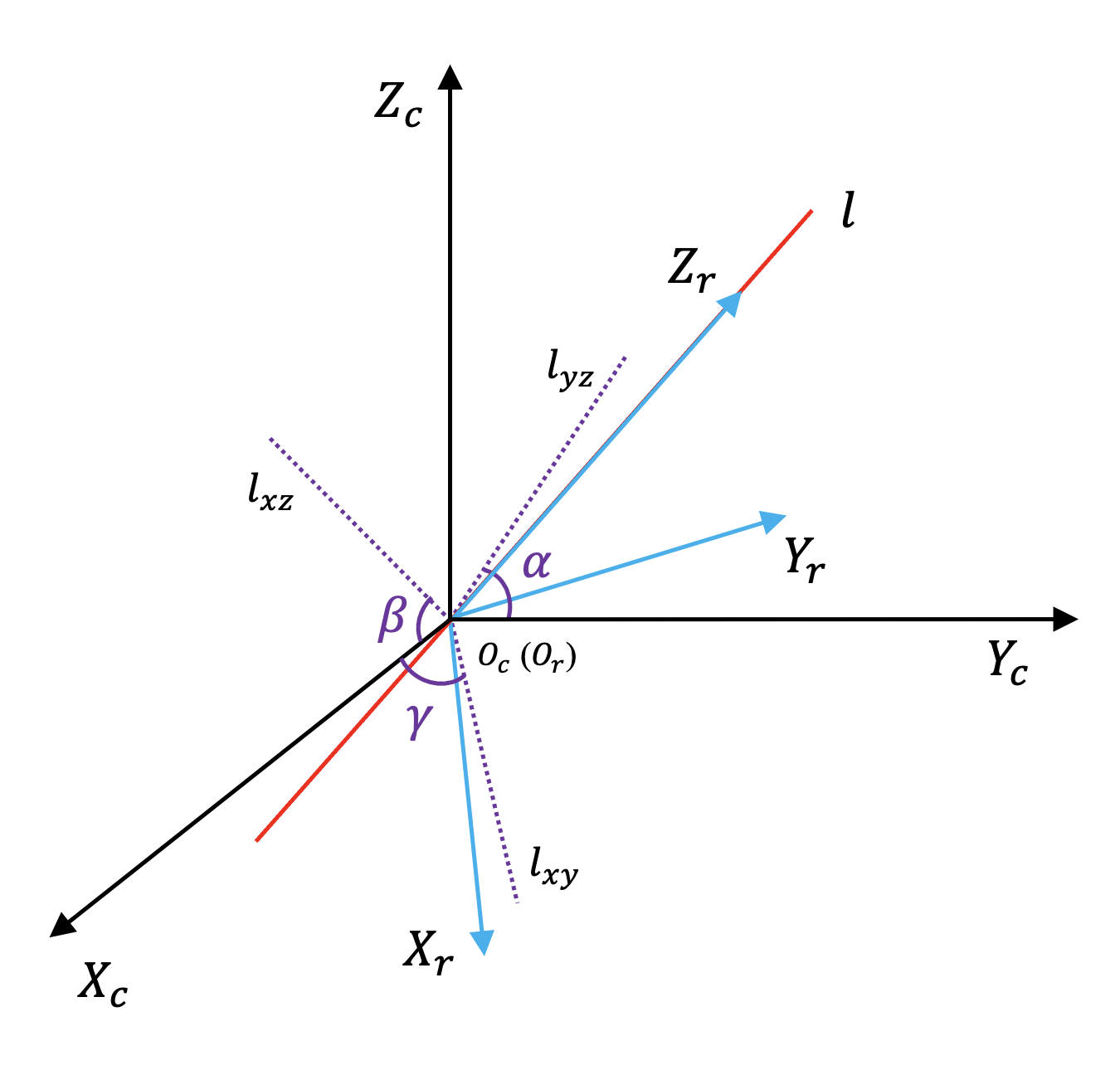}
\caption{Cartesian coordinate system $O_{\rm c}$--$X_{\rm c}$$Y_{\rm
    c}$$Z_{\rm c}$ (black) centered at the OC centre. Also shown is
  the rotational Cartesian coordinate system $O_{\rm r}$--$X_{\rm
    r}$$Y_{\rm r}$$Z_{\rm r}$ (blue), where $\vec{l}$ is the
  rotation axis, in accordance with the $Z_{\rm r}$--axis.
  The projections of $\vec{l}$ in the $Y_{\rm c}$--$Z_{\rm c}$, the $X_{\rm
    c}$--$Z_{\rm c}$, and the $X_{\rm c}$--$Y_{\rm c}$ plane are
  $l_{yz}$, $l_{xz}$ and $l_{xy}$, respectively.
  $\alpha$, $\beta$ and $\gamma$ indicate the included angles between
  $l_{yz}$ and the $Y_{\rm c}$--axis, $l_{xz}$ and the $X_{\rm
    c}$--axis, $l_{xy}$ and the $X_{\rm c}$--axis, respectively. The
  projection of the $X_{\rm r}$--axis in the $X_{\rm c}$--$Y_{\rm c}$
  plane is set to $l_{xy}$.
}
\label{fig:ccs2axis}
\end{center}
\end{figure}

\section{Methods}
\label{methods}

With the celestial positions, parallaxes, proper motions, and radial
velocities given by {\it Gaia} EDR3,
we can determine the 3D spatial coordinates ($x_{\rm g}$, $y_{\rm g}$,
$z_{\rm g}$) and the 3D velocity components ($v_{x_{\rm g}}$,
$v_{y_{\rm g}}$, $v_{z_{\rm g}}$) of the cluster
members~\citep[][]{xu2013,reid2019}. 
The calculations are in the
Galactic Cartesian coordinate system ($O_{\rm g}$--$X_{\rm g}$$Y_{\rm
g}$$Z_{\rm g}$). The origin of coordinates $O_{\rm g}$ is in the
Galactic centre.
In this step, the distances of sources are derived by inverting
the trigonometric parallaxes. The position and velocity
uncertainties are estimated with a Monte Carlo method by taking 
into account the observation errors of astrometric parameters 
but without considering the covariances between parallaxes and 
proper motions.
At the location of the Sun, the $X_{\rm g}$--axis points to the
Galactic centre, the $Y_{\rm g}$--axis towards the direction of the
Galactic rotation, and the $Z_{\rm g}$--axis towards the north
Galactic pole, respectively.
The circular rotation speed at the Sun's position is adopted as
236 $\pm$ 7~km s$^{-1}$, and the Sun is at a distance of 8.15 
$\pm$ 0.15 kpc to the Galactic centre~\citep{reid2019}.
The three velocity components of the solar motion, i.e., towards the
Galactic centre ($U_{\odot}$), in the direction of Galactic rotation
($V_{\odot}$), towards the north Galactic North Pole ($W_{\odot}$), are
adopted as ($U_{\odot}$, $V_{\odot}$, $W_{\odot}$)~=~(10.6 $\pm$ 1.2,
10.7 $\pm$ 6.0, 7.6 $\pm$ 0.7)~km~s$^{-1}$~\citep{reid2019}.

By defining the OC centre as the origin of coordinates, a new
Cartesian coordinate system $O_{\rm c}$--$X_{\rm c}$$Y_{\rm c}$$Z_{\rm
c}$ can be established.
The OC centre and its uncertainty in the Galactic coordinate
system come from the means and standard deviations of the astrometric
parameters of the selected member stars by performing the Monte Carlo 
simulations.
The 3D spatial coordinates ($x_{\rm c}$, $y_{\rm c}$, $z_{\rm c}$) and
the 3D velocity components ($v_{x_{\rm c}}$, $v_{y_{\rm c}}$,
$v_{z_{\rm c}}$) of a member star in this coordinate system can be
calculated as:
%

\begin{equation}
  \begin{pmatrix}
   x_{\rm c} \\ y_{\rm c} \\ z_{\rm c}
  \end{pmatrix} =
  \begin{pmatrix}
   x_{\rm g} \\ y_{\rm g} \\ z_{\rm g}
  \end{pmatrix} - 
    \begin{pmatrix}
   x_{\rm g, s} \\ y_{\rm g, s} \\ z_{\rm g, s}
  \end{pmatrix}, 
  \label{equ:gc2ccs}
 \end{equation}
\begin{equation}
    \begin{pmatrix}
   v_{x_{\rm c}} \\ v_{y_{\rm c}} \\ v_{z_{\rm c}}
  \end{pmatrix} =
  \begin{pmatrix}
   v_{x_{\rm g}} \\ v_{y_{\rm g}} \\ v_{z_{\rm g}} 
  \end{pmatrix} - 
  \begin{pmatrix}
   v_{x_{\rm g, s}} \\ v_{y_{\rm g, s}} \\ v_{z_{\rm g, s}}
  \end{pmatrix},
  \label{equ:gc2ccsv}
 \end{equation}
where ($x_{\rm g, s}$, $y_{\rm g, s}$, $z_{\rm g, s}$) are the
coordinates of the cluster centre, ($v_{x_{\rm g, s}}$, $v_{y_{\rm g,
s}}$, $v_{y_{\rm g, s}}$) represent the 3D systemic velocities of
the cluster in the Galactic Cartesian coordinate system.
The vectors ($v_{x_{\rm c}}$, $v_{y_{\rm c}}$, $v_{z_{\rm c}}$) are
therefore the 3D residual velocities of a member star.

As shown in Figure~\ref{fig:ccs2axis}, if a cluster does rotate, the
rotation axis $\vec{l}$ of the cluster in the $O_{\rm c}$--$X_{\rm c}$$Y_{\rm
c}$$Z_{\rm c}$ system can be determined from the three position
angles (PA), $\alpha$, $\beta$, and $\gamma$.
Here, the vector $\vec{l}$ across the origin $O_{\rm c}$ has a 
positive direction above the $X_{\rm c}$--$Y_{\rm c}$ plane.
Based on the observational data, these angles can be fitted by a
residual velocity method adopted in many previous
studies~\citep[e.g.,][]{bellazzini2012,
lanzoni2013,ferraro2018,lanzoni2018,loktin2020,leanza2022}.
In the following, we describe the procedure to determine the angle
$\alpha$:

\begin{enumerate}
\item The projection of the rotation axis $\vec{l}$ on the $Y_{\rm
  c}$--$Z_{\rm c}$ plane, $l_{yz}$, will divide the cluster members
  into two sub-samples (Figure~\ref{fig:ccs2axis}). If the cluster
  does rotate, the mean residual velocities $v_{x_{\rm c}}$ for the
  member stars of the two different sub-samples should present
  opposite signs.
\item Staring from the $Y_{\rm c}$--axis ($\alpha$~$=0^\circ$), the
  projection of the rotation axis on the $Y_{\rm c}$--$Z_{\rm c}$
  plane, $l_{yz}$ rotates counterclockwise as the increase of
  $\alpha$. A series of the mean residual velocities $v_{x_{\rm c}}$
  for the member stars of the two different sub-samples can be
  calculated. Meanwhile, the Monte Carlo simulations can be
  preformed to randomly simulate the mean residual velocities and
  the uncertainties based on the $v_{x_{\rm c}}$ values as well as
  their uncertainties of the member stars.
\item Then, we can determine the dependence of the mean residual 
  velocities on the PA. The mean residual velocity 
  reaches a maximum/minimum at a PA $\alpha$, which 
  indicates the projection of the actual rotation axis $\vec{l}$ of the cluster 
  on the $Y_{\rm c}$--$Z_{\rm c}$ plane.
\end{enumerate}

Similar methods are used to determine the values of $\beta$ and
$\gamma$.
As shown in Figure~\ref{fig:ccs2axis}, $\alpha$, $\beta$, and $\gamma$
satisfy the relation: tan $\alpha$ $\cdot$ tan $\gamma$ = tan $\beta$,
which can be used as a test of the three angles calculated by the
above method.
%
\setlength{\tabcolsep}{5.0mm}
\begin{table}[ht]
\centering
\caption{Definitions of the included angle $\alpha_{i}$, $\beta_{i}$
  and $\gamma_{i}$ between the axes of the Cartesian coordinate system
  $O_{\rm c}$--$X_{\rm c}$$Y_{\rm c}$$Z_{\rm c}$ centered at the OC
  centre and those of the rotational Cartesian coordinate system
  $O_{\rm r}$--$X_{\rm r}$$Y_{\rm r}$$Z_{\rm r}$ (also see
  Figure~\ref{fig:ccs2axis}).}
\begin{tabular}{c|ccc} 
\hline
                          & $X_{\rm c}$-axis & $Y_{\rm c}$-axis & $Z_{\rm c}$-axis  \\ \hline 
 $X_{\rm r}$-axis &     $\alpha_{1}$     &  $\beta_{1}$       &   $\gamma_{1}$     \\  
 $Y_{\rm r}$-axis &     $\alpha_{2}$     &  $\beta_{2}$       &   $\gamma_{2}$     \\  
 $Z_{\rm r}$-axis &     $\alpha_{3}$     &  $\beta_{3}$       &   $\gamma_{3}$     \\    \hline  
 \end{tabular}
 \label{table:table1}
\end{table}

Once the rotation axis $\vec{l}$ of the cluster is determined, we can
construct a rotational Cartesian coordinate system ($O_{\rm
  r}$--$X_{\rm r}$$Y_{\rm r}$$Z_{\rm r}$) for the cluster, where the
origin of coordinates $O_{\rm r}$ is located at the cluster centre.
The $Z_{\rm r}$--axis is in accordance with the rotation axis $\vec{l}$. The
projection of the $X_{\rm r}$--axis in the $X_{\rm c}$--$Y_{\rm c}$
plane is set to be consistent with the projection $l_{xy}$ of $\vec{l}$
(Figure~\ref{fig:ccs2axis}).
Therefore, the vectors of $X_{\rm r}$--axis, $Y_{\rm r}$--axis, and
$Z_{\rm r}$--axis in the Cartesian coordinate system $O_{\rm
  c}$--$X_{\rm c}$$Y_{\rm c}$$Z_{\rm c}$ can be determined by the
angles of $\alpha$, $\beta$, and $\gamma$, i.e.,

\begin{equation}
\left\{
\begin{aligned}
& \overrightarrow{O_{\rm r}X_{\rm r}} = (1, {\rm tan}~\gamma, -~\frac{1 + {\rm tan}^2\gamma}{{\rm tan}~\beta}), \\
& \overrightarrow{O_{\rm r}Y_{\rm r}} = (-~{\rm tan}~\gamma, 1, 0), \\
& \overrightarrow{O_{\rm r}Z_{\rm r}} = (\frac{1}{{\rm tan}~\beta}, \frac{1}{{\rm tan}~\alpha}, 1).
 \end{aligned}
\right.
\label{equ:vecrcc}
\end{equation}

The 3D coordinates and 3D velocity components in the $O_{\rm
  c}$--$X_{\rm c}$$Y_{\rm c}$$Z_{\rm c}$ system are related to those
in the $O_{\rm r}$--$X_{\rm r}$$Y_{\rm r}$$Z_{\rm r}$ system by:

\begin{equation}
  \begin{pmatrix}
   x_{\rm r} \\ y_{\rm r} \\ z_{\rm r}
  \end{pmatrix} = 
  \begin{pmatrix}
{\rm cos}~\alpha_{1} & {\rm cos}~\beta_{1} & {\rm cos}~\gamma_{1} \\ {\rm cos}~\alpha_{2} & {\rm cos}~\beta_{2} & {\rm cos}~\gamma_{2} \\ {\rm cos}~\alpha_{3} & {\rm cos}~\beta_{3} & {\rm cos}~\gamma_{3}
  \end{pmatrix}
  \begin{pmatrix}
   x_{\rm c} \\ y_{\rm c} \\ z_{\rm c}
  \end{pmatrix},
  \label{equ:ccs2rcc}
\end{equation}
\begin{equation}
  \begin{pmatrix}
   v_{x_{\rm r}} \\ v_{y_{\rm r}} \\ v_{z_{\rm r}}
  \end{pmatrix} = 
  \begin{pmatrix}
{\rm cos}~\alpha_{1} & {\rm cos}~\beta_{1} & {\rm cos}~\gamma_{1} \\ {\rm cos}~\alpha_{2} & {\rm cos}~\beta_{2} & {\rm cos}~\gamma_{2} \\ {\rm cos}~\alpha_{3} & {\rm cos}~\beta_{3} & {\rm cos}~\gamma_{3}
  \end{pmatrix}
  \begin{pmatrix}
   v_{x_{\rm c}} \\ v_{y_{\rm c}} \\ v_{z_{\rm c}}
  \end{pmatrix},
  \label{equ:ccs2rccv}
 \end{equation}
here, $\alpha_{i}$, $\beta_{i}$ and $\gamma_{i}$ are the included
angles listed in Table~\ref{table:table1}.

The cylindrical coordinate system ($r$, $\varphi$, $z$) is more
convenient to study the rotational properties of stellar
clusters~\citep[e.g.,][]{lanzoni2018}, which is adopted in the
following analysis.
The transformation equations from the Cartesian coordinates 
to the cylindrical coordinates are:
\begin{equation}
\left\{
\begin{aligned}
&  r =  \sqrt{x_{\rm r}^{2} + y_{\rm r}^{2}}, \\
& \varphi =  {\rm atan}~\frac{y_{\rm r}}{x_{\rm r}}, \\
&  z =  z_{\rm r}.
 \end{aligned}
\right.
\label{equ:rcc2cc}
\end{equation}
\begin{equation}
  \begin{pmatrix}
   v_{r} \\
   v_{\varphi} \\
   v_{z}
  \end{pmatrix} = 
  \begin{pmatrix}
  {\rm cos}~\varphi  & {\rm sin}~\varphi & 0 \\
  -{\rm sin}~\varphi & {\rm cos}~\varphi & 0 \\
  0 & 0 & 1
  \end{pmatrix}
  \begin{pmatrix}
   v_{x_{\rm r}} \\
   v_{y_{\rm r}} \\
   v_{z_{\rm r}}
  \end{pmatrix}.
  \label{equ:rcc2ccv}
\end{equation}
%
%

\begin{figure*}
\centering
\includegraphics[scale=0.15]{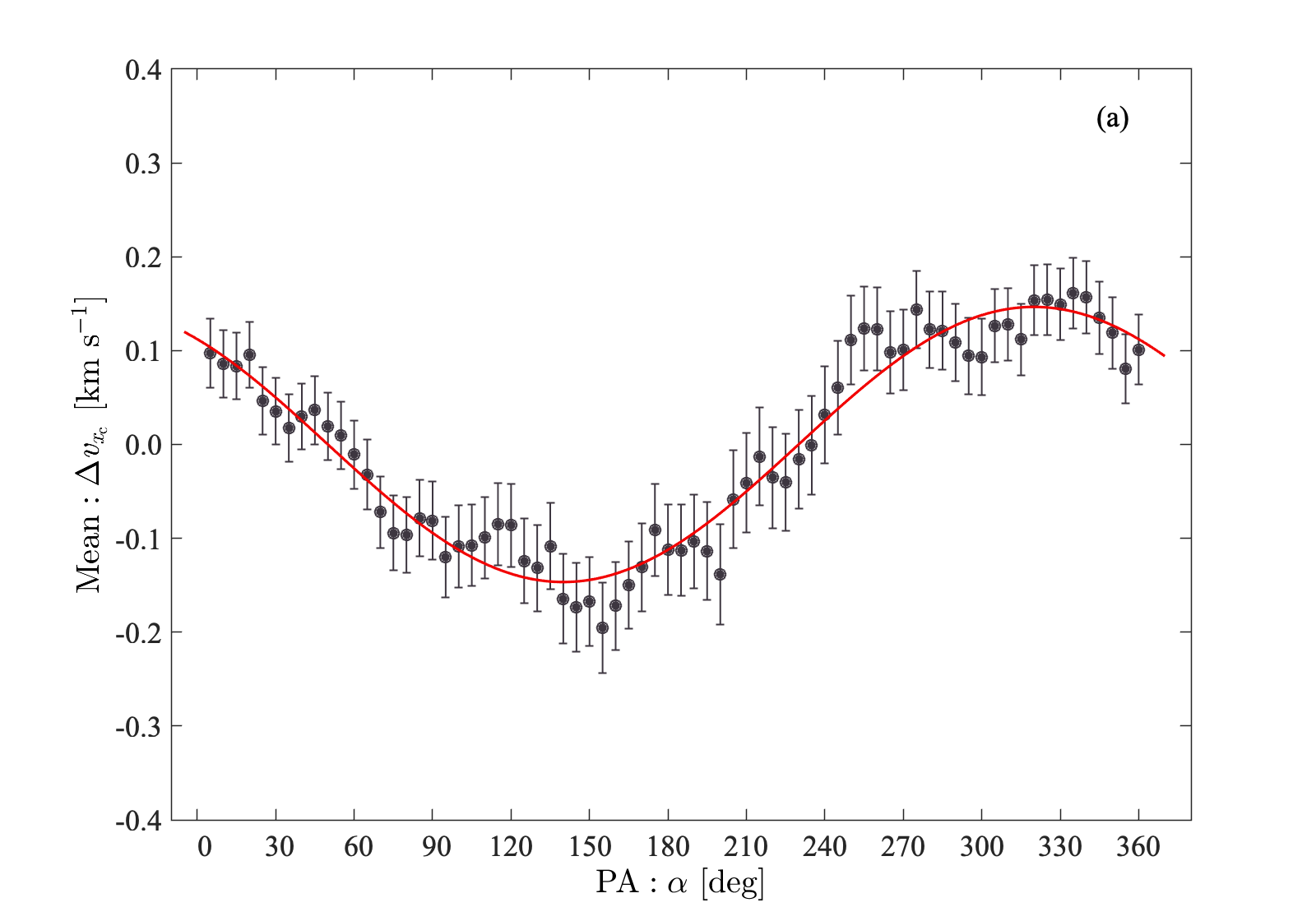}
\includegraphics[scale=0.15]{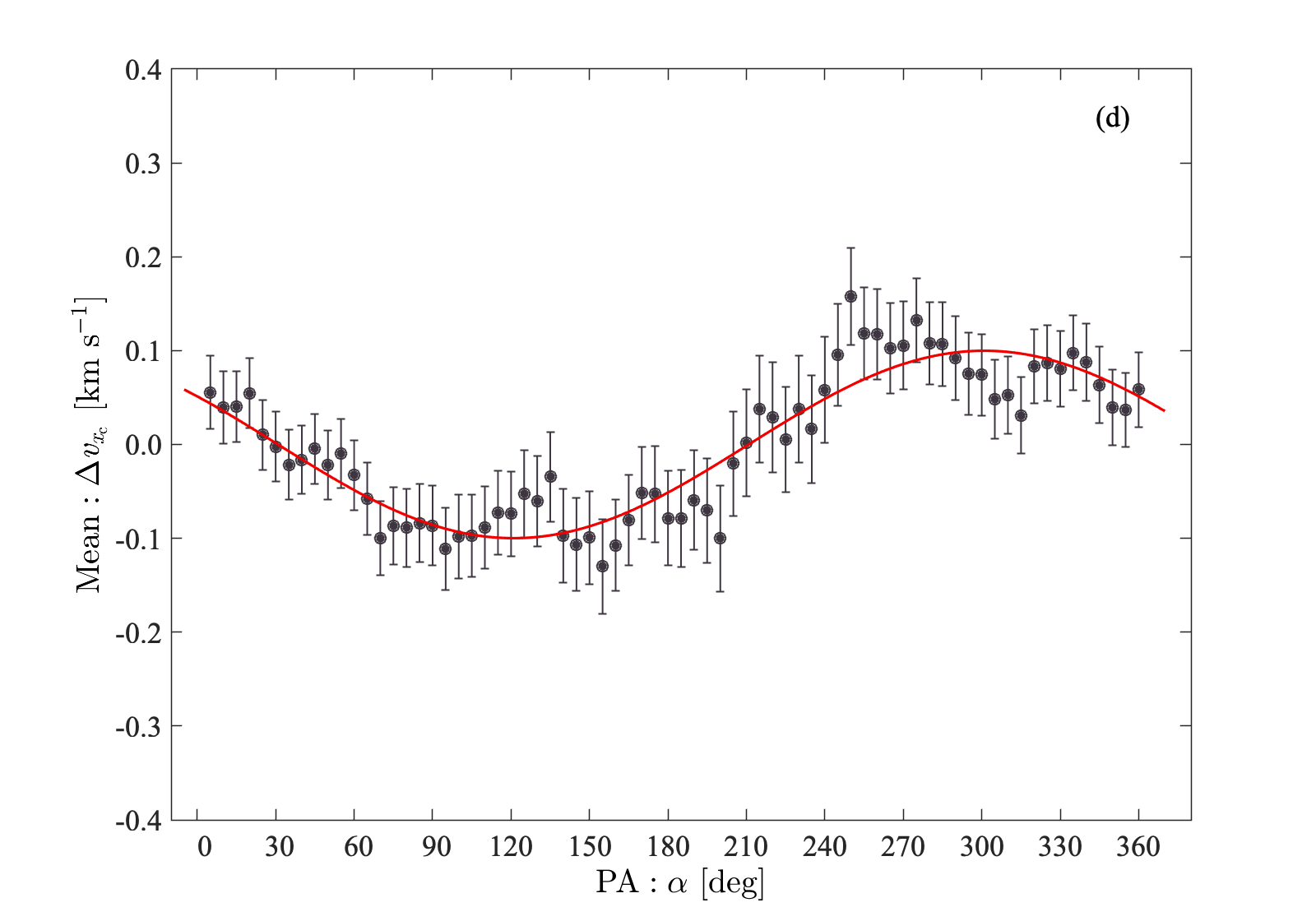}
\includegraphics[scale=0.15]{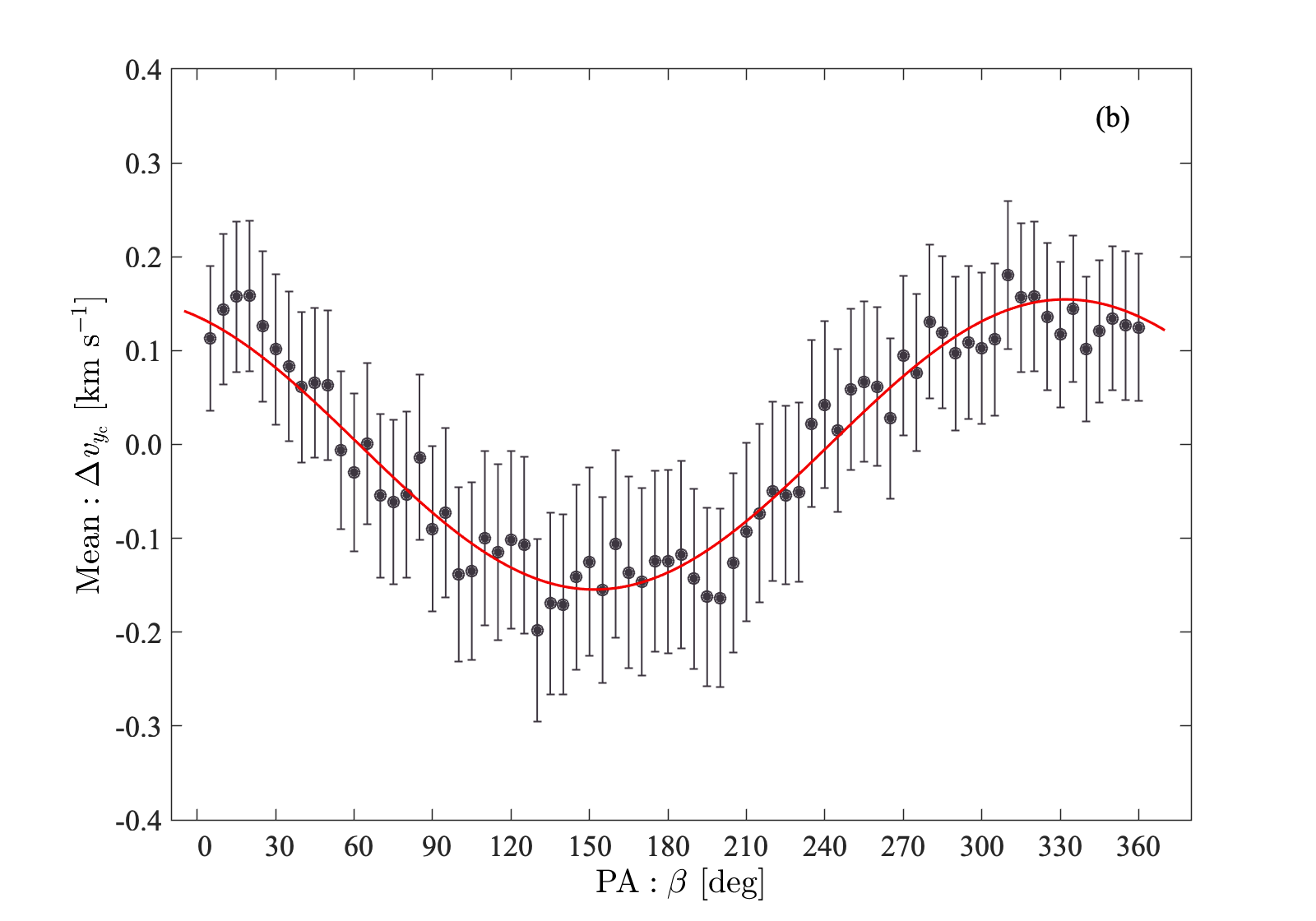}
\includegraphics[scale=0.15]{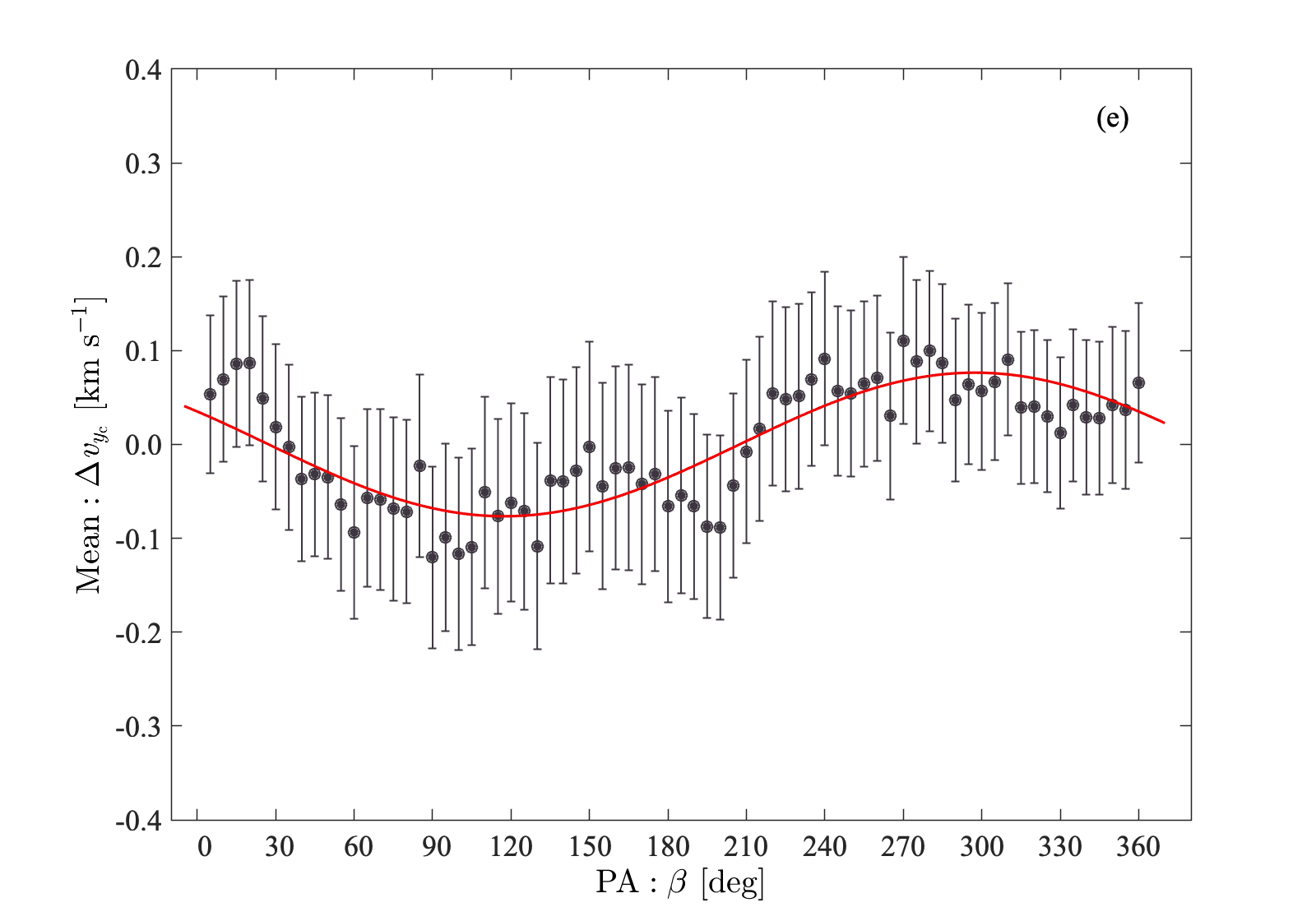}
\includegraphics[scale=0.15]{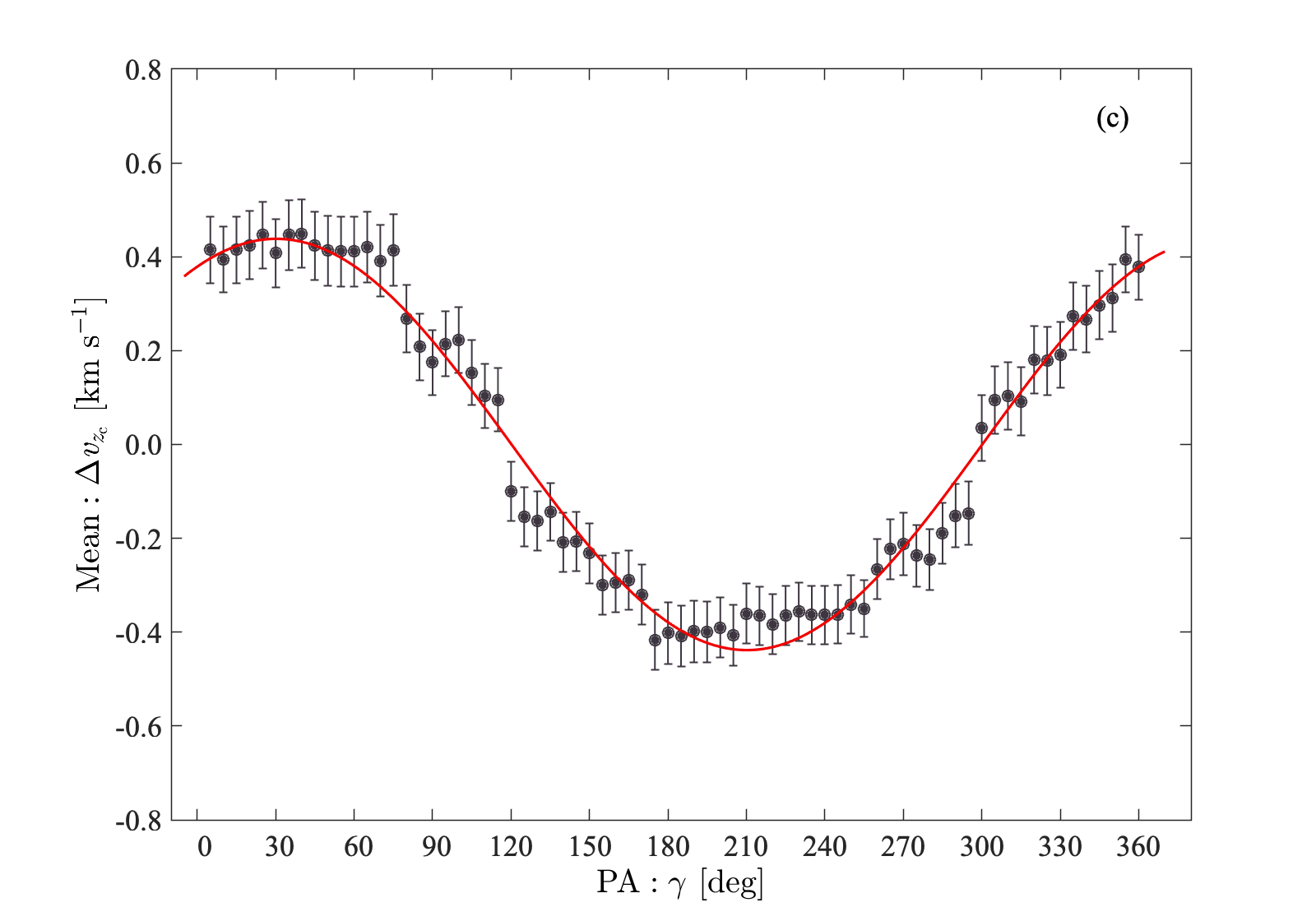}
\includegraphics[scale=0.15]{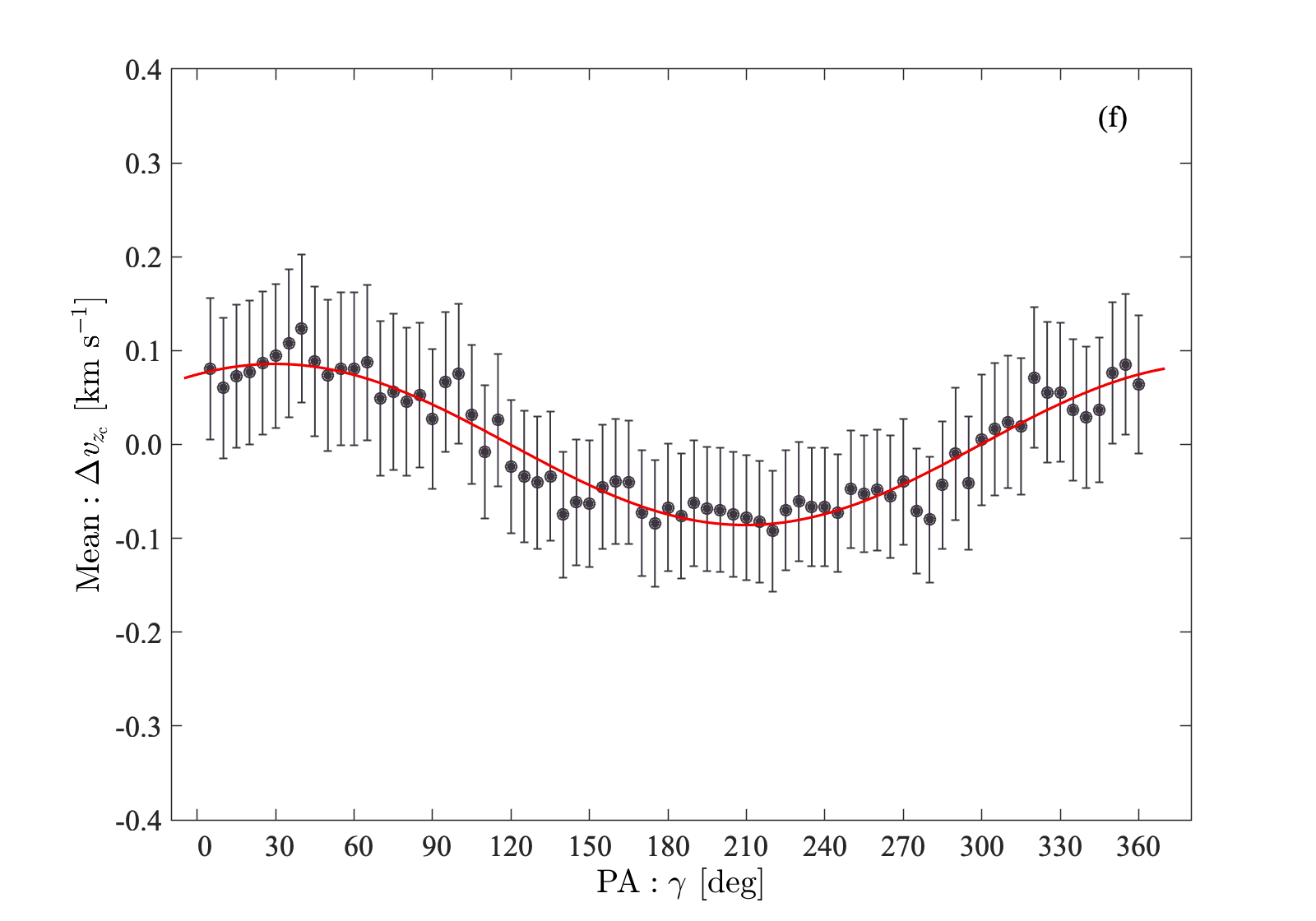}
\caption{Mean residual velocities of $v_{x_{\rm c}}$, $v_{y_{\rm
        c}}$ and $v_{z_{\rm c}}$ as a function of the position angle
    (PA) are shown in {\it Panel} (a), (b), and (c), respectively, for
    all the member stars. The results given in {\it Panel} (d), (e),
    and (f) are similar to that of the left column, but derived from
    the member stars within the tidal radius of Praesepe. 
    The error bars (gray), as well as the best-fitting sine functions (red) are
  also shown in the plots. }
\label{fig:PA}
\end{figure*}

\section{Results}\label{sec:results}

\subsection{Rotation in Praesepe}\label{sec:axis}

According to the astrometric parameters of the 172 member 
stars (see Sect.~\ref{sample}), the means and standard deviations 
of the fundamental parameters of Praesepe are determined as
($Ra$, $Dec$) =
($129.85^{\circ}$ $\pm$ $1.51^{\circ}$, $19.52^{\circ}$ $\pm$
$1.49^{\circ}$), $\varpi$ = $5.41$ $\pm$ $0.20$ mas, proper 
motions ($\mu_{\alpha^{*}}$, $\mu_{\delta}$) = ($-35.82$ $\pm$ 
$1.54$, $-12.79$ $\pm$ $1.09$) mas yr$^{-1}$, and RV = 
$35.1$ $\pm$ $1.4$ km s$^{-1}$, which are in good agreement 
with previous determinations~\citep[e.g.,][]{babusiaux2018,
roser2019,lodieu2019,gao2019,loktin2020}.
Here, these parameters do not take into account the 
measurement errors as weights.
After applying the Monte Carlo simulation to the means and 
standard deviations of the parameters from the selected member 
stars with 3D kinematic measurements, we obtain the centre, systemic 
motion, and the corresponding uncertainties of Praesepe in the 
Galactic coordinate system.
For each of the member stars, we first calculate its 3D coordinates,
3D velocities, as well as the uncertainties in the Galactic
Cartesian coordinate system, then transform them to the $O_{\rm
  c}$--$X_{\rm c}$$Y_{\rm c}$$Z_{\rm c}$ system according to
Equ.~(\ref{equ:gc2ccs}) and Equ.~(\ref{equ:gc2ccsv}).
To investigate the rotation and estimate the rotation axis as well as
velocity of the Praesepe cluster, we calculate the mean residual 
velocity of the member stars as a function of the position angle 
following the method described in Sect.~\ref{methods}. A Monte 
Carlo method is used to estimate the uncertainties. A step length 
of $5^{\circ}$ for the position angle is adopted in the calculation.
The mean residual velocities for the first half of the member stars
correspond to the PA from $5^{\circ}$ to $180^{\circ}$, and that 
of the second half correspond to the PA from $180^{\circ}$ to
$360^{\circ}$.
Figure~\ref{fig:PA}(a), (b), and (c) present the mean residual
velocity of all the member stars as a function of the PA for
$v_{x_{\rm c}}$, $v_{y_{\rm c}}$, and $v_{z_{\rm c}}$,
respectively. The results for those member stars within the tidal
radius are shown in Figure~\ref{fig:PA}(d), (e), and (f) for
comparison. 
They all present sinusoidal behaviors, which indicate the rotation of
the Praesepe's system.
Here the adopted tidal radius of Praesepe is 10~pc.
The best-fitting position angles $\alpha$, $\beta$, and $\gamma$ 
with all the member stars in the sample
are $\alpha$ = $139.9^{\circ}$ $\pm$ $2.9^{\circ}$,
$\beta$ = $152.2^{\circ}$ $\pm$ $6.4^{\circ}$, and $\gamma$ =
$210.2^{\circ}$ $\pm$ $1.5^{\circ}$, respectively.
In comparison, the obtained position angles based on the member
stars within the tidal radius are $\alpha$ = $120.8^{\circ}$ $\pm$
$4.5^{\circ}$, $\beta$ = $118.1^{\circ}$ $\pm$ $16.5^{\circ}$, and
$\gamma$ = $211.2^{\circ}$ $\pm$ $7.8^{\circ}$, respectively.
These fitted position angles satisfy the relation of tan $\alpha$
$\cdot$ tan $\gamma$ = tan $\beta$ considering the uncertainties.

The member stars beyond the tidal radius are no longer simply
controlled by the cluster itself, but
influenced by the gravitational force of the Galaxy. Actually, most
of the OCs can cross the Galactic plane several times in one orbital
period of them~\citep[e.g.,][]{wu2009}.
Taking this into account, we adopt the best-fitting position
angles from the member stars within the cluster tidal radius in
the following analysis. Considering the relation of tan
$\alpha$ $\cdot$ tan $\gamma$ = tan $\beta$, the position angles of
($\alpha$, $\beta$, $\gamma$) = ($120.8^{\circ}$, $134.5^{\circ}$,
$211.2^{\circ}$) are adopted in this study to derive the ($r$,
$\varphi$, $z$) and ($v_{r}$, $v_{\varphi}$, $v_{z}$) of the member
stars in the cylindrical coordinate system. The
corresponding uncertainties are estimated by a Monte Carlo method.
The angle between the rotation axis of the Praesepe cluster
and the Galactic plane is estimated to be $41^\circ\pm12^\circ$.
Meanwhile, based on the rotational velocities of the member
stars, the mean rotational velocity of Praesepe within its tidal
radius is estimated to be 
0.2 $\pm$ 0.05~km~s$^{-1}$,
which is concordant with the properties show in Figure~\ref{fig:PA}(d), (e), and (f).
Here, the error bar is the uncertainty of the mean rotational velocity,
obtained through the Monte Carlo simulation.

Generally, dense molecular cores are the nurseries of embedded
clusters, which are predecessors of OCs~\citep[e.g.,][]{lada2003}.
Hydrodynamical simulations show that the stellar component of the
embedded cluster can inherit the rotation signature from the parent
gas and the rotation is common for embedded
clusters~\citep{mapelli2017}. The rotation of Praesepe
probably originates from the rotational characteristic of its
predecessor embedded cluster. From the derived rotational velocities
of member stars, it is shown that not all the member stars rotate in
the same direction, although this may be partially influenced by the
astrometric measurement uncertainties. Such phenomenon also occurs
in the rotation of global
clusters~\citep[e.g.,][]{lanzoni2018,leanza2022}. OCs are the
survivors of hierarchical or substructured protoclusters that
gestated in molecular clouds. Many stellar feedback mechanisms play
important roles in the formation of protoclusters, such as
protostellar outflows, stellar radiation pressure, stellar winds
from massive stars, etc. The mutual interference during stellar
cluster formation may result in that not all the member stars of
Praesepe neatly rotate in the same direction.
%

\begin{figure}
\centering
\includegraphics[scale=0.17]{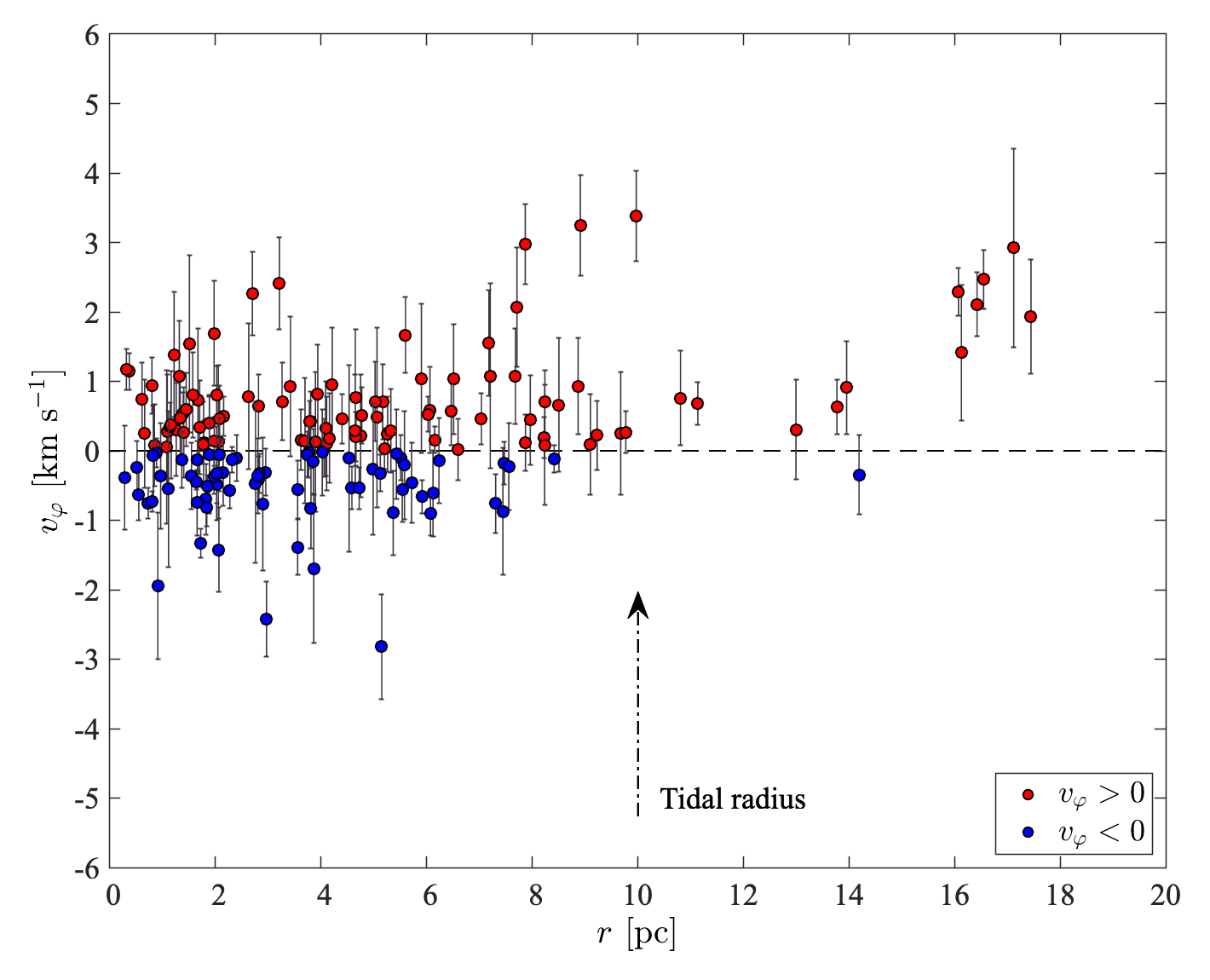}
\caption{The rotational velocity components as a function of the
  distance $r$ from the cluster centre, for the cluster members within
  two times the tidal radius of Praesepe.}
\label{fig:r_vphi}
\end{figure}

\subsection{Rotational properties of stars in Praesepe}
\label{sec:rotation}

Figure~\ref{fig:r_vphi} shows the rotational velocities $v_{\varphi}$
of the cluster members as a function of the distance $r$ from the
cluster centre. A Monte Carlo method is used to estimate the
uncertainties.
The referenced value of the tidal radius of Praesepe is adopted as
10~pc.
Inside the tidal radius (10~pc) of the Praesepe cluster, we find that
the member stars in the inner region have slightly larger rotational
velocities than those in the outer region.
Near or beyond the tidal radius of Praesepe, the rotational velocities
of the member stars tend to present large dispersion, which may be 
due to that these stars are partially influenced by the Galactic tidal
force, rather than simply dominated by the gravitational force of the
cluster itself.
Inside the the tidal radius, there are several stars with peculiarly
rotational velocities that deviate significantly from the main part,
which may be ``passing through'' stars and not the bona fide members
of the Praesepe cluster.
The kinematic property of stars in an OC, e.g., whether the
rotation of member stars follow the classical Newton's theorems, is
an interesting question that has not been well addressed. This issue
can be explored by comparing the observational results with the
theoretical expectations of Newton's theorems.
In the following, we primarily analyze the properties of $v_{\varphi}$
of the member stars within the tidal radius of Praesepe.
Firstly, assuming that the OC system follows a spherically symmetric 
density distribution, then the gravitational potential at the radius $r$ is:
\begin{flalign}
\begin{split}
\Phi (r) = - {\rm 4 \pi} G \ [\frac{1}{r} \int_{0}^{r}\rho(r') r'^{2} \ {\rm d}r' +  \int_{r}^{\infty}\rho(r') r' \ {\rm d}r'].
\label{equ:phi}
\end{split}
\end{flalign}
The asymmetry assumption of the data distribution is evaluated by
calculating the skewness of the sample. The skewness of any
perfectly symmetric distribution is zero. If the skewness is between
$-$0.50 and 0.50, the data are suggested to be fairly symmetrical. The
Pearson's moment coefficients of skewness of the Galactic longitude,
latitude, and parallaxes of all the member stars within the tidal
radius of Praesepe are 0.01, $-$0.13, and 0.11, respectively. The
results indicate that the stars in the parent catalog described in
Sect.~\ref{sample} present a well spherically symmetric
distribution. For the selected 172 member stars with 3D kinematic
measurements, the corresponding skewness coefficients are 0.28,
$-$0.35, and 0.15, respectively, which are slightly larger than
those of the parent sample but still show a good spherically
symmetric distribution.

From the Newton's theorems, the gravitational attraction of the system
at $r$ is entirely determined by the mass interior to $r$, i.e.,
\begin{flalign}
\begin{split}
& {\rm \bf{F}} (r) = -\frac{{\rm d} \Phi}{{\rm d} r} \ \hat{e}_{r} = - \frac{GM(r)}{r^{2}} \ \hat{e}_{r}, \\
& M (r) = {\rm 4 \pi} \int_{0}^{r}\rho(r') r'^{2} \ {\rm d}r'.
\label{equ:f}
\end{split}
\end{flalign}
Here, $M (r)$ is the mass inside the radius $r$, $G$ = $4.3 \times
10^{-3}~{\rm pc~M}_{\odot}^{-1}~({\rm km~s}^{-1})^{2}$ is the
gravitational constant.
The circular speed $v_{c} (r)$ of the system therefore can be
calculated by:
\begin{flalign}
\begin{split}
v_{c}^{2} (r) = r \frac{{\rm d} \Phi}{{\rm d} r} = r | {\rm \bf{F}} | = \frac{GM(r)}{r}.
\label{equ:vc}
\end{split}
\end{flalign}
Here, this formula is used in an attempt to describe the rotational 
property of Praesepe, since its rotation has been identified.
Although individual stars may not present perfectly circular motion,  
we can try to make an approximate comparison between the root-mean-square (RMS) 
rotational velocity derived from cluster members and the theoretical 
expectation with Equ.~(\ref{equ:vc}).
Actually, many of the member stars may have residual radial
motions as discussed in the next subsection.
The same as \cite{roser2019}, the mass density of Praesepe adopted in
this study is described by a Plummer model~\citep{plummer1915}, which
is:
\begin{flalign}
\begin{split}
\rho(r) = \frac{3M_{\rm t}}{4\pi r_{\rm co}^{3}}~\frac{1}{[1~+~(r/r_{\rm co})^{2}]^{5/2}}.
\label{equ:rho}
\end{split}
\end{flalign}
Here, $M_{\rm t}$ is the total mass inside the tidal radius of
Praesepe, and $r_{\rm co}$ is the core radius.
Then, $M (r)$ and the corresponding circular speed $v_{c}
(r)$ can be expressed as:
\begin{flalign}
\begin{split}
M(r) = M_{\rm t} \cdot \frac{r^{3}}{(r^2~+~r_{\rm co}^{2})^{3/2}},
\label{equ:mass-r}
\end{split}
\end{flalign}
\begin{flalign}
\begin{split}
v_{\rm c} (r) = [G M_{\rm t} \cdot \frac{r^{2}}{(r^2~+~r_{\rm co}^{2})^{3/2}}]^{1/2}.
\label{equ:v-t}
\end{split}
\end{flalign}
%

\begin{figure}
\centering
\includegraphics[scale=0.17]{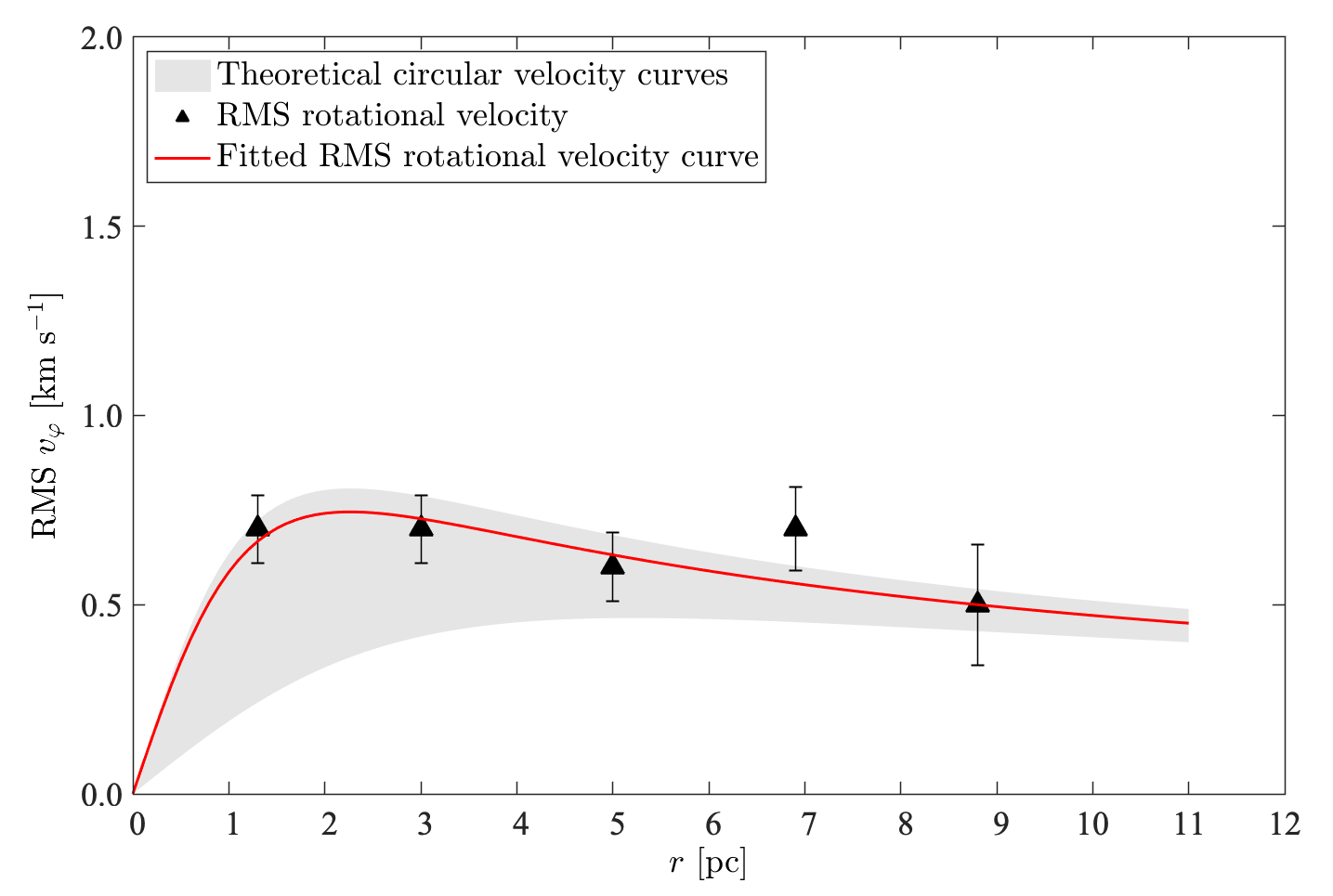}
\caption{Comparison between the observed RMS rotational
    velocities of the member stars (black triangles) and the
    theoretical values at distance $r$ from the cluster centre. The
    grey shadow shows the theoretical values calculated by
    Equ.~(\ref{equ:v-t}) after considering the possible ranges of the
    tidal mass and the core radius of Praesepe. The red line is a
    best-fitting curve to the observed RMS rotational velocities.}
\label{fig:v_t}
\end{figure}

\setlength{\tabcolsep}{3.2mm}
\begin{table}[ht]
\centering 
\caption{RMS rotational velocities of the members stars within
    the tidal radius of Praesepe, which are derived from the
    observational data of {\it Gaia} EDR3.}
\begin{tabular}{cc|cccc} 
  \hline
$R_{i}$  &  $R_{o}$ & $R_{m}$  & $N$  & RMS $v_{\varphi}$  &  $\epsilon_{{\rm RMS} \, v_{\varphi}}$   \\ 
      {[pc]}  &  {[pc]}    &  {[pc]}  &  & {[km~s$^{-1}$]} & {[km~s$^{-1}$]}  \\
       (1)    &   (2)   &  (3)     & (4)  & (5) & (6) \\
\hline
 0.0 &  2.0    &  1.3  &  46  &  0.7  &  0.1   \\  
 2.0 &  4.0    &  3.0  &  35  &  0.7  &  0.1   \\  
 4.0 &  6.0    &  5.0  &  32  &  0.6  &  0.1   \\  
 6.0 &  8.0    &  6.9  &  19  &  0.7  &  0.1   \\  
 8.0 & 10.0    &  8.8  &  10  &  0.5  &  0.2   \\    \hline  
 \end{tabular}
 \tablecomments{Columns 1--2: the inner and outer radius of each bin;
   Columns 3--6: the average radius, number of the member stars,
     RMS rotational velocity and its uncertainty derived from the
     stars in each bin.}
 \label{table:table2}
\end{table}

The tidal mass $M_{\rm t}$ and the core radius $r_{\rm co}$ of
Praesepe have been determined by many research works.
The estimated value of $M_{\rm t}$ is in the range of [483, 630]~${\rm
  M}_{\odot}$
\citep[e.g.,][]{roser2019,wang2014,kraus2007,gao2019,adams2002,holland2000}.
The core radius $r_{\rm co}$ is in the range of [1.6, 3.7] 
pc~\citep[e.g.,][]{mermilliod1990,adams2002,lodieu2019,
roser2019,gao2019}.
The difference between the theoretical circular velocities derived from 
different tidal masses is not very significant, less than 0.1 km s$^{-1}$.
While the maximum difference for the derived core radius is $\sim$ 0.3
km s$^{-1}$.
According to the ranges of $M_{\rm t}$ and $r_{\rm co}$, the possible
theoretical circular velocity curves are calculated and shown in
Figure~\ref{fig:v_t}.

For the member stars inside the tidal radius of Praesepe, the
median absolute value of their rotational velocities is $\sim$
0.5~km~s$^{-1}$, and the standard deviation of $v_{\varphi}$ is
$\sim$ 1.0 km s$^{-1}$. We notice that there are eight stars 
with $|v_{\varphi}|$ larger than 2.0~km s$^{-1}$, and deviate from those
of the vast majority of the member stars.
In order to reduce the influence of the stars with
peculiar rotational velocities and investigate the features
indicated by vast majority of the member stars, the cluster members
with values of $|v_{\varphi}|$ smaller than 2.0 km s$^{-1}$ are
extracted from the sample and divided into several bins.
Then, we calculate the RMS rotational velocity of the member
stars in each bin. The Monte Carlo simulations are used to estimate
the uncertainties. The results are shown in
Table~\ref{table:table2} and Figure~\ref{fig:v_t}.
The derived RMS rotational velocities are in agreement with
the theoretical values.
By fitting the RMS velocities listed in Table~\ref{table:table2} with
Equ.~(\ref{equ:v-t}), we obtain a core radius of 1.6 $\pm$ 0.5~pc
and a tidal mass of 537 $\pm$ 146~${\rm M}_{\odot}$ for the 
Praesepe cluster.
The uncertainties are estimated by the Monte Carlo method.
The best-fitting RMS velocity curve shown in Figure~\ref{fig:v_t} also
indicates that the rotation of member stars within the tidal radius of 
Praesepe probably follow the Newton's theorems.
Besides, according to the best-fitting curve, the member stars at the 
periphery of the Praesepe cluster may have rotational velocities of about 
0.4 km s$^{-1}$. 
%
%

\begin{figure}
\centering
\includegraphics[scale=0.17]{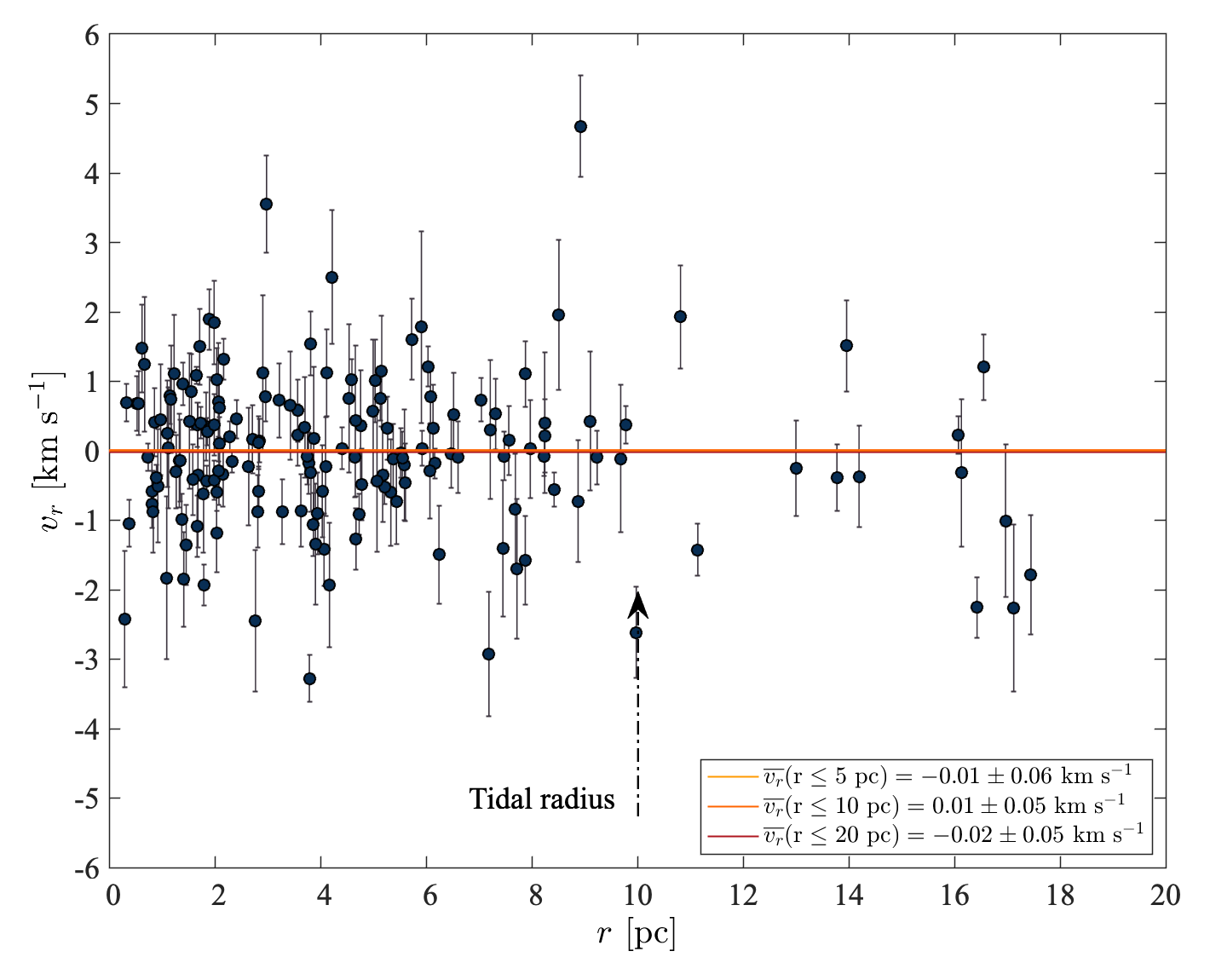}
\caption{The radial velocity components and uncertainties as a function of the
  distance $r$ from the cluster centre, for the cluster members within
  two times the tidal radius of Praesepe. The mean values of the
  radial components within 5~pc, 10~pc, and 20~pc of the cluster
  centre all are close to zero, and shown in the panel.}
\label{fig:r_vr}
\end{figure}

\subsection{Absence of expansion or contraction}
\label{sec:ec}

During the emerging of an OC from its natal cloud, it can expand for 
a long period before reaching an equilibrium
state~\cite[e.g.,][]{kroupa2001,lada2003,banerjee2017}.
The age of Praesepe is about 590–660 Myr
~\citep[e.g.,][]{mermilliod1981,vandenberg1984,delorme2011,
brandt2015,gossage2018}. It is not sure whether the Praesepe cluster 
is still expanding.

The expansion or contraction of Praesepe can be understood by the
statistical analysis on the radial component $v_{r}$ of the member
stars perpendicular to the rotation axis.
Figure~\ref{fig:r_vr} shows the radial components of the cluster
members within two times the tidal radius (10~pc) of Praesepe as a
function of the distance $r$ from the cluster centre, which does not
present a visible indication of expansion or contraction.
In addition, the mean radial component $\overline{v_{r}}$ of the
member stars within 5~pc, 10~pc, and 20~pc of the cluster centre are
$-$0.01 km s$^{-1}$, 0.01 km s$^{-1}$, and $-$0.02 km s$^{-1}$, 
with uncertainties of 0.06 km s$^{-1}$, 0.05 km s$^{-1}$, and 0.05 
km s$^{-1}$,
respectively. Here, the error bars are the uncertainties in the mean, 
obtained through Monte Carlo simulations.
The mean radial components are close to zero show that there is no 
significant indication of expansion or contraction for the Praesepe cluster, 
implying that the rotation of member stars within the tidal radius
of Praesepe can present the closed-loop motion.
Figure~\ref{fig:r_vr} also implies that there will be a
non-negligible radial velocity acceleration and a radial velocity
dispersion contribution to the support of the cluster, although the
radial velocity dispersion is partially influenced by the
astrometric measurement uncertainties. The absence of expansion or
contraction of Praesepe suggests that the cluster system should
possess additional mass to provide force for supporting the radial
velocity acceleration of member stars. It is speculated that the
derived dynamic mass in Sect.~\ref{sec:rotation} is a lower limit
for the Praesepe cluster. 

\section{Summary}
\label{summary}

In this work, we explored the kinematic properties of Praesepe, 
the only OC in the Milky Way whose rotation has been 
investigated exclusively.
Based on the high-precision astrometric dataset of {\it Gaia} 
EDR3, the rotation in the Praesepe cluster and its rotation axis 
in the Galaxy were determined by analysing the 3D residual 
velocities of cluster members for the first time.
Our developed methodology also derived the rotational velocity 
components of the member stars, suggesting that the member 
stars within the tidal radius of the Praesepe cluster probably 
also conforms to the theorems of Newtonian mechanics.
Additionally, the current results suggest no significant indication 
of expansion or contraction for Praesepe.

\acknowledgments We thank the anonymous referee for the instructive 
comments and suggestions which help us a lot to improve the paper. 
This work was funded by the NSFC, grant numbers 11933011, 11873019 and
11673066, and by the Key Laboratory for Radio Astronomy. YJL thanks
supports from the Natural Science Foundation of Jiangsu Province
(grant number BK20210999) and the Entrepreneurship and Innovation
Program of Jiangsu Province. LGH thanks the support from the
  Youth Innovation Promotion Association CAS. We used data from the
European Space Agency mission \textit{Gaia}
(\url{http://www.cosmos.esa.int/gaia}), processed by the \textit{Gaia}
Data Processing and Analysis Consortium (DPAC; see
\url{http://www.cosmos.esa.int/web/gaia/dpac/consortium}). Funding for
DPAC has been provided by national institutions, in particular the
institutions participating in the \textit{Gaia} Multilateral
Agreement.

\appendix

\setcounter{table}{0}
\setcounter{figure}{0}
\renewcommand{\thetable}{A\arabic{table}}
\renewcommand{\thefigure}{A\arabic{figure}}

Table~\ref{table:table_s1} presents the astrometric parameters and errors of examples of selected member 
stars with 3D kinematic measurements in this work.
%
\begin{table*}
\centering
\caption{Astrometric parameters and errors of selected member stars with 3D kinematic measurements.}
\scriptsize
\setlength{\tabcolsep}{2.0mm}
\renewcommand\arraystretch{1.0}
\begin{tabular}{l  c  c c c r@{ $\pm$ }l  r@{ $\pm$ }l  l@{ $\pm$ }r  r  c  r@{ $\pm$ }l }
\hline \hline \\
\multicolumn{1}{c}{$Gaia$ ID} &
\multicolumn{1}{c}{$\alpha$} &
\multicolumn{1}{c}{$\delta$} &
\multicolumn{1}{c}{$l$} &
\multicolumn{1}{c}{$b$} &
\multicolumn{2}{c}{$\varpi$} &
\multicolumn{2}{c}{$\mu_{\alpha^{*}}$} &
\multicolumn{2}{c}{$\mu_{\delta}$} &
\multicolumn{1}{c}{$G$} &
\multicolumn{1}{c}{bp$-$rp} &
\multicolumn{2}{c}{$V_{r}$}
\\
\multicolumn{1}{c}{} &
\multicolumn{1}{c}{[deg]} &
\multicolumn{1}{c}{[deg]} &
\multicolumn{1}{c}{[deg]} &
\multicolumn{1}{c}{[deg]} &
\multicolumn{2}{c}{[mas]} &
\multicolumn{2}{c}{[mas yr$^{-1}$]} &
\multicolumn{2}{c}{[mas yr$^{-1}$]} &
\multicolumn{1}{c}{[mag]} &
\multicolumn{1}{c}{[mag]} &
\multicolumn{2}{c}{[km s$^{-1}$]}
 \\ \\
 \hline \\
  660944717521395840 & 131.07 &  18.74 & 207.34 &  33.02 &   5.43 &   0.02 & $-$35.98 &   0.02 & $-$11.63 &   0.02 &  12.07 &   1.06 &  33.66 &   0.71 \\
  660954067666386944 & 131.32 &  18.89 & 207.27 &  33.30 &   5.54 &   0.02 & $-$37.62 &   0.02 & $-$12.54 &   0.02 &   9.52 &   0.57 &  34.54 &   0.54 \\
  660957980380034688 & 130.95 &  18.80 & 207.22 &  32.93 &   5.40 &   0.02 & $-$36.14 &   0.02 & $-$11.78 &   0.01 &   9.98 &   0.64 &  36.67 &   1.89 \\
  660962863759344768 & 130.98 &  18.89 & 207.12 &  32.99 &   5.42 &   0.02 & $-$36.34 &   0.02 & $-$12.80 &   0.01 &  12.39 &   1.15 &  33.96 &   0.85 \\
  660998975844267264 & 130.20 &  18.90 & 206.80 &  32.30 &   5.34 &   0.05 & $-$36.69 &   0.04 & $-$12.92 &   0.03 &  11.27 &   0.89 &  35.87 &   0.40 \\
  661004400386393984 & 130.34 &  18.93 & 206.83 &  32.44 &   5.40 &   0.02 & $-$36.58 &   0.02 & $-$12.45 &   0.01 &  12.65 &   1.22 &  34.72 &   0.89 \\
  661015915195312512 & 130.78 &  19.07 & 206.85 &  32.88 &   5.50 &   0.05 & $-$36.69 &   0.05 & $-$12.01 &   0.04 &   9.44 &   0.67 &  33.15 &   0.69 \\
  661018904492612224 & 130.93 &  19.08 & 206.90 &  33.01 &   5.07 &   0.02 & $-$33.69 &   0.02 & $-$12.72 &   0.01 &  13.04 &   1.33 &  37.21 &   1.18 \\
  661028662658249344 & 130.68 &  19.10 & 206.78 &  32.80 &   5.57 &   0.03 & $-$37.39 &   0.03 & $-$13.66 &   0.02 &  11.44 &   0.92 &  34.84 &   0.37 \\
  661029384212747392 & 130.76 &  19.17 & 206.73 &  32.89 &   5.39 &   0.02 & $-$36.68 &   0.02 & $-$11.38 &   0.02 &  11.91 &   1.02 &  34.00 &   0.64 \\
  661029727808613760 & 130.67 &  19.13 & 206.74 &  32.80 &   5.30 &   0.02 & $-$35.65 &   0.02 & $-$11.61 &   0.02 &  11.92 &   1.01 &  35.55 &   0.57 \\
  661031480156796544 & 130.58 &  19.15 & 206.68 &  32.73 &   5.36 &   0.01 & $-$36.07 &   0.02 & $-$11.65 &   0.01 &  13.09 &   1.40 &  35.14 &   1.27 \\
  661111538347104512 & 131.70 &  19.64 & 206.56 &  33.89 &   5.41 &   0.02 & $-$36.79 &   0.02 & $-$12.68 &   0.01 &  10.59 &   0.75 &  33.59 &   0.70 \\
  661206577385260800 & 129.99 &  19.20 & 206.39 &  32.23 &   5.46 &   0.02 & $-$36.70 &   0.02 & $-$12.75 &   0.01 &   9.38 &   0.56 &  35.76 &   0.42 \\
  661207024061875456 & 129.80 &  19.12 & 206.41 &  32.03 &   5.42 &   0.02 & $-$35.45 &   0.02 & $-$12.26 &   0.01 &  10.45 &   0.73 &  35.10 &   0.99 \\
  661211727047722752 & 130.11 &  19.22 & 206.42 &  32.34 &   5.45 &   0.02 & $-$36.58 &   0.03 & $-$11.35 &   0.02 &  12.56 &   1.23 &  34.96 &   1.71 \\
  661212933936844416 & 130.11 &  19.28 & 206.36 &  32.36 &   5.22 &   0.03 & $-$35.30 &   0.03 & $-$11.53 &   0.02 &  11.01 &   0.82 &  33.31 &   1.07 \\
  661216369910684544 & 130.03 &  19.31 & 206.29 &  32.29 &   5.44 &   0.04 & $-$36.05 &   0.04 & $-$11.89 &   0.03 &  10.81 &   0.93 &  36.81 &   1.09 \\
  661221764391246080 & 130.44 &  19.27 & 206.50 &  32.65 &   5.59 &   0.02 & $-$36.92 &   0.02 & $-$12.99 &   0.02 &  10.05 &   0.67 &  35.13 &   0.31 \\
  661222279785743616 & 130.55 &  19.28 & 206.53 &  32.74 &   5.44 &   0.02 & $-$36.82 &   0.02 & $-$12.38 &   0.01 &  11.93 &   1.03 &  34.29 &   0.47 \\
  661234473194513664 & 130.45 &  19.46 & 206.29 &  32.72 &   5.62 &   0.05 & $-$38.55 &   0.06 & $-$10.85 &   0.04 &  13.22 &   1.46 &  37.46 &   1.35 \\
  661235989321322240 & 130.67 &  19.54 & 206.28 &  32.94 &   5.37 &   0.02 & $-$36.98 &   0.02 & $-$12.38 &   0.01 &   9.67 &   0.59 &  36.12 &   1.88 \\
  661239390935360512 & 130.28 &  19.45 & 206.23 &  32.57 &   5.51 &   0.02 & $-$38.02 &   0.02 & $-$12.07 &   0.01 &  12.17 &   1.09 &  35.39 &   0.79 \\
  661242895630246016 & 130.22 &  19.48 & 206.17 &  32.53 &   5.38 &   0.02 & -35.34 &   0.02 & -12.31 &   0.02 &  10.18 &   0.67 &  34.43 &   0.32 \\
  661243273585807872 & 130.09 &  19.46 & 206.14 &  32.41 &   5.41 &   0.02 & -35.37 &   0.02 & -12.81 &   0.02 &  10.54 &   0.74 &  36.76 &   0.49 \\
  \multicolumn{13}{c}{\hspace{3.3cm}  ... } 
 \\ \\  \hline
\end{tabular}
\tablecomments{The full table is available online.}
 \label{table:table_s1}
\end{table*}



\bibliography{sample63}{}

\begin{thebibliography}{}
\expandafter\ifx\csname natexlab\endcsname\relax\def\natexlab#1{#1}\fi
\providecommand{\url}[1]{\href{#1}{#1}}
\providecommand{\dodoi}[1]{doi:~\href{http://doi.org/#1}{\nolinkurl{#1}}}
\providecommand{\doeprint}[1]{\href{http://ascl.net/#1}{\nolinkurl{http://ascl.net/#1}}}
\providecommand{\doarXiv}[1]{\href{https://arxiv.org/abs/#1}{\nolinkurl{https://arxiv.org/abs/#1}}}

\bibitem[{{Adams} {et~al.}(2002){Adams}, {Stauffer}, {Skrutskie}, {Monet},
  {Portegies Zwart}, {Janes}, \& {Beichman}}]{adams2002}
{Adams}, J.~D., {Stauffer}, J.~R., {Skrutskie}, M.~F., {et~al.} 2002, \aj, 124,
  1570, \dodoi{10.1086/342016}

\bibitem[{{Banerjee} \& {Kroupa}(2017)}]{banerjee2017}
{Banerjee}, S., \& {Kroupa}, P. 2017, \aap, 597, A28,
  \dodoi{10.1051/0004-6361/201526928}

\bibitem[{{Barnes}(2007)}]{Barnes2007}
{Barnes}, S.~A. 2007, \apj, 669, 1167, \dodoi{10.1086/519295}

\bibitem[{{Bellazzini} {et~al.}(2012){Bellazzini}, {Bragaglia}, {Carretta},
  {Gratton}, {Lucatello}, {Catanzaro}, \& {Leone}}]{bellazzini2012}
{Bellazzini}, M., {Bragaglia}, A., {Carretta}, E., {et~al.} 2012, \aap, 538,
  A18, \dodoi{10.1051/0004-6361/201118056}

\bibitem[{{Bertelli Motta} {et~al.}(2017){Bertelli Motta}, {Salaris},
  {Pasquali}, \& {Grebel}}]{Motta2017}
{Bertelli Motta}, C., {Salaris}, M., {Pasquali}, A., \& {Grebel}, E.~K. 2017,
  \mnras, 466, 2161, \dodoi{10.1093/mnras/stw3252}

\bibitem[{{Brandt} \& {Huang}(2015)}]{brandt2015}
{Brandt}, T.~D., \& {Huang}, C.~X. 2015, \apj, 807, 24,
  \dodoi{10.1088/0004-637X/807/1/24}

\bibitem[{{Cantat-Gaudin} {et~al.}(2018){Cantat-Gaudin}, {Jordi}, {Vallenari},
  {Bragaglia}, {Balaguer-N{\'u}{\~n}ez}, {Soubiran}, {Bossini}, {Moitinho},
  {Castro-Ginard}, {Krone-Martins}, {Casamiquela}, {Sordo}, \&
  {Carrera}}]{cantat2018}
{Cantat-Gaudin}, T., {Jordi}, C., {Vallenari}, A., {et~al.} 2018, \aap, 618,
  A93, \dodoi{10.1051/0004-6361/201833476}

\bibitem[{{Cantat-Gaudin} {et~al.}(2020){Cantat-Gaudin}, {Anders},
  {Castro-Ginard}, {Jordi}, {Romero-G{\'o}mez}, {Soubiran}, {Casamiquela},
  {Tarricq}, {Moitinho}, {Vallenari}, {Bragaglia}, {Krone-Martins}, \&
  {Kounkel}}]{cantat2020}
{Cantat-Gaudin}, T., {Anders}, F., {Castro-Ginard}, A., {et~al.} 2020, \aap,
  640, A1, \dodoi{10.1051/0004-6361/202038192}

\bibitem[{{Castro-Ginard} {et~al.}(2020){Castro-Ginard}, {Jordi}, {Luri},
  {{\'A}lvarez Cid-Fuentes}, {Casamiquela}, {Anders}, {Cantat-Gaudin},
  {Mongui{\'o}}, {Balaguer-N{\'u}{\~n}ez}, {Sol{\`a}}, \& {Badia}}]{castro2020}
{Castro-Ginard}, A., {Jordi}, C., {Luri}, X., {et~al.} 2020, \aap, 635, A45,
  \dodoi{10.1051/0004-6361/201937386}

\bibitem[{{Castro-Ginard} {et~al.}(2021){Castro-Ginard}, {McMillan}, {Luri},
  {Jordi}, {Romero-G{\'o}mez}, {Cantat-Gaudin}, {Casamiquela}, {Tarricq},
  {Soubiran}, \& {Anders}}]{castro2021}
{Castro-Ginard}, A., {McMillan}, P.~J., {Luri}, X., {et~al.} 2021, \aap, 652,
  A162, \dodoi{10.1051/0004-6361/202039751}

\bibitem[{{Castro-Ginard} {et~al.}(2022){Castro-Ginard}, {Jordi}, {Luri},
  {Cantat-Gaudin}, {Carrasco}, {Casamiquela}, {Anders},
  {Balaguer-N{\'u}{\~n}ez}, \& {Badia}}]{castro2022}
{Castro-Ginard}, A., {Jordi}, C., {Luri}, X., {et~al.} 2022, \aap, 661, A118,
  \dodoi{10.1051/0004-6361/202142568}

\bibitem[{{Delorme} {et~al.}(2011){Delorme}, {Collier Cameron}, {Hebb},
  {Rostron}, {Lister}, {Norton}, {Pollacco}, \& {West}}]{delorme2011}
{Delorme}, P., {Collier Cameron}, A., {Hebb}, L., {et~al.} 2011, \mnras, 413,
  2218, \dodoi{10.1111/j.1365-2966.2011.18299.x}

\bibitem[{{Dias} {et~al.}(2002){Dias}, {Alessi}, {Moitinho}, \&
  {L{\'e}pine}}]{dias2002}
{Dias}, W.~S., {Alessi}, B.~S., {Moitinho}, A., \& {L{\'e}pine}, J.~R.~D. 2002,
  \aap, 389, 871, \dodoi{10.1051/0004-6361:20020668}

\bibitem[{{Dobbie} {et~al.}(2004){Dobbie}, {Pinfield}, {Napiwotzki}, {Hambly},
  {Burleigh}, {Barstow}, {Jameson}, \& {Hubeny}}]{dobbie2004}
{Dobbie}, P.~D., {Pinfield}, D.~J., {Napiwotzki}, R., {et~al.} 2004, \mnras,
  355, L39, \dodoi{10.1111/j.1365-2966.2004.08522.x}

\bibitem[{{Dobbie} {et~al.}(2006){Dobbie}, {Napiwotzki}, {Burleigh}, {Barstow},
  {Boyce}, {Casewell}, {Jameson}, {Hubeny}, \& {Fontaine}}]{dobbie2006}
{Dobbie}, P.~D., {Napiwotzki}, R., {Burleigh}, M.~R., {et~al.} 2006, \mnras,
  369, 383, \dodoi{10.1111/j.1365-2966.2006.10311.x}

\bibitem[{{Ferraro} {et~al.}(2018){Ferraro}, {Mucciarelli}, {Lanzoni},
  {Pallanca}, {Lapenna}, {Origlia}, {Dalessandro}, {Valenti}, {Beccari},
  {Bellazzini}, {Vesperini}, {Varri}, \& {Sollima}}]{ferraro2018}
{Ferraro}, F.~R., {Mucciarelli}, A., {Lanzoni}, B., {et~al.} 2018, \apj, 860,
  50, \dodoi{10.3847/1538-4357/aabe2f}

\bibitem[{{Gaia Collaboration} {et~al.}(2016){Gaia Collaboration}, {Prusti},
  {de Bruijne}, {Brown}, {Vallenari}, {Babusiaux}, {Bailer-Jones}, {Bastian},
  {Biermann}, {Evans}, {Eyer}, {Jansen}, {Jordi}, {Klioner}, {Lammers},
  {Lindegren}, {Luri}, {Mignard}, {Milligan}, {Panem}, {Poinsignon},
  {Pourbaix}, {Randich}, {Sarri}, {Sartoretti}, {Siddiqui}, {Soubiran},
  {Valette}, {van Leeuwen}, {Walton}, {Aerts}, {Arenou}, {Cropper}, {Drimmel},
  {H{\o}g}, {Katz}, {Lattanzi}, {O'Mullane}, {Grebel}, {Holland}, {Huc},
  {Passot}, {Bramante}, {Cacciari}, {Casta{\~n}eda}, {Chaoul}, {Cheek}, {De
  Angeli}, {Fabricius}, {Guerra}, {Hern{\'a}ndez}, {Jean-Antoine-Piccolo},
  {Masana}, {Messineo}, {Mowlavi}, {Nienartowicz}, {Ord{\'o}{\~n}ez-Blanco},
  {Panuzzo}, {Portell}, {Richards}, {Riello}, {Seabroke}, {Tanga},
  {Th{\'e}venin}, {Torra}, {Els}, {Gracia-Abril}, {Comoretto},
  {Garcia-Reinaldos}, {Lock}, {Mercier}, {Altmann}, {Andrae}, {Astraatmadja},
  {Bellas-Velidis}, {Benson}, {Berthier}, {Blomme}, {Busso}, {Carry},
  {Cellino}, {Clementini}, {Cowell}, {Creevey}, {Cuypers}, {Davidson}, {De
  Ridder}, {de Torres}, {Delchambre}, {Dell'Oro}, {Ducourant}, {Fr{\'e}mat},
  {Garc{\'\i}a-Torres}, {Gosset}, {Halbwachs}, {Hambly}, {Harrison}, {Hauser},
  {Hestroffer}, {Hodgkin}, {Huckle}, {Hutton}, {Jasniewicz}, {Jordan},
  {Kontizas}, {Korn}, {Lanzafame}, {Manteiga}, {Moitinho}, {Muinonen},
  {Osinde}, {Pancino}, {Pauwels}, {Petit}, {Recio-Blanco}, {Robin}, {Sarro},
  {Siopis}, {Smith}, {Smith}, {Sozzetti}, {Thuillot}, {van Reeven}, {Viala},
  {Abbas}, {Abreu Aramburu}, {Accart}, {Aguado}, {Allan}, {Allasia},
  {Altavilla}, {{\'A}lvarez}, {Alves}, {Anderson}, {Andrei}, {Anglada Varela},
  {Antiche}, {Antoja}, {Ant{\'o}n}, {Arcay}, {Atzei}, {Ayache}, {Bach},
  {Baker}, {Balaguer-N{\'u}{\~n}ez}, {Barache}, {Barata}, {Barbier}, {Barblan},
  {Baroni}, {Barrado y Navascu{\'e}s}, {Barros}, {Barstow}, {Becciani},
  {Bellazzini}, {Bellei}, {Bello Garc{\'\i}a}, {Belokurov}, {Bendjoya},
  {Berihuete}, {Bianchi}, {Bienaym{\'e}}, {Billebaud}, {Blagorodnova},
  {Blanco-Cuaresma}, {Boch}, {Bombrun}, {Borrachero}, {Bouquillon}, {Bourda},
  {Bouy}, {Bragaglia}, {Breddels}, {Brouillet}, {Br{\"u}semeister},
  {Bucciarelli}, {Budnik}, {Burgess}, {Burgon}, {Burlacu}, {Busonero}, {Buzzi},
  {Caffau}, {Cambras}, {Campbell}, {Cancelliere}, {Cantat-Gaudin}, {Carlucci},
  {Carrasco}, {Castellani}, {Charlot}, {Charnas}, {Charvet}, {Chassat},
  {Chiavassa}, {Clotet}, {Cocozza}, {Collins}, {Collins}, {Costigan}, {Crifo},
  {Cross}, {Crosta}, {Crowley}, {Dafonte}, {Damerdji}, {Dapergolas}, {David},
  {David}, {De Cat}, {de Felice}, {de Laverny}, {De Luise}, {De March}, {de
  Martino}, {de Souza}, {Debosscher}, {del Pozo}, {Delbo}, {Delgado},
  {Delgado}, {di Marco}, {Di Matteo}, {Diakite}, {Distefano}, {Dolding}, {Dos
  Anjos}, {Drazinos}, {Dur{\'a}n}, {Dzigan}, {Ecale}, {Edvardsson}, {Enke},
  {Erdmann}, {Escolar}, {Espina}, {Evans}, {Eynard Bontemps}, {Fabre},
  {Fabrizio}, {Faigler}, {Falc{\~a}o}, {Farr{\`a}s Casas}, {Faye}, {Federici},
  {Fedorets}, {Fern{\'a}ndez-Hern{\'a}ndez}, {Fernique}, {Fienga}, {Figueras},
  {Filippi}, {Findeisen}, {Fonti}, {Fouesneau}, {Fraile}, {Fraser}, {Fuchs},
  {Furnell}, {Gai}, {Galleti}, {Galluccio}, {Garabato}, {Garc{\'\i}a-Sedano},
  {Gar{\'e}}, {Garofalo}, {Garralda}, {Gavras}, {Gerssen}, {Geyer}, {Gilmore},
  {Girona}, {Giuffrida}, {Gomes}, {Gonz{\'a}lez-Marcos},
  {Gonz{\'a}lez-N{\'u}{\~n}ez}, {Gonz{\'a}lez-Vidal}, {Granvik}, {Guerrier},
  {Guillout}, {Guiraud}, {G{\'u}rpide}, {Guti{\'e}rrez-S{\'a}nchez}, {Guy},
  {Haigron}, {Hatzidimitriou}, {Haywood}, {Heiter}, {Helmi}, {Hobbs},
  {Hofmann}, {Holl}, {Holland}, {Hunt}, {Hypki}, {Icardi}, {Irwin}, {Jevardat
  de Fombelle}, {Jofr{\'e}}, {Jonker}, {Jorissen}, {Julbe}, {Karampelas},
  {Kochoska}, {Kohley}, {Kolenberg}, {Kontizas}, {Koposov}, {Kordopatis},
  {Koubsky}, {Kowalczyk}, {Krone-Martins}, {Kudryashova}, {Kull}, {Bachchan},
  {Lacoste-Seris}, {Lanza}, {Lavigne}, {Le Poncin-Lafitte}, {Lebreton},
  {Lebzelter}, {Leccia}, {Leclerc}, {Lecoeur-Taibi}, {Lemaitre}, {Lenhardt},
  {Leroux}, {Liao}, {Licata}, {Lindstr{\o}m}, {Lister}, {Livanou}, {Lobel},
  {L{\"o}ffler}, {L{\'o}pez}, {Lopez-Lozano}, {Lorenz}, {Loureiro},
  {MacDonald}, {Magalh{\~a}es Fernandes}, {Managau}, {Mann}, {Mantelet},
  {Marchal}, {Marchant}, {Marconi}, {Marie}, {Marinoni}, {Marrese},
  {Marschalk{\'o}}, {Marshall}, {Mart{\'\i}n-Fleitas}, {Martino}, {Mary},
  {Matijevi{\v{c}}}, {Mazeh}, {McMillan}, {Messina}, {Mestre}, {Michalik},
  {Millar}, {Miranda}, {Molina}, {Molinaro}, {Molinaro}, {Moln{\'a}r},
  {Moniez}, {Montegriffo}, {Monteiro}, {Mor}, {Mora}, {Morbidelli}, {Morel},
  {Morgenthaler}, {Morley}, {Morris}, {Mulone}, {Muraveva}, {Musella},
  {Narbonne}, {Nelemans}, {Nicastro}, {Noval}, {Ord{\'e}novic},
  {Ordieres-Mer{\'e}}, {Osborne}, {Pagani}, {Pagano}, {Pailler}, {Palacin},
  {Palaversa}, {Parsons}, {Paulsen}, {Pecoraro}, {Pedrosa}, {Pentik{\"a}inen},
  {Pereira}, {Pichon}, {Piersimoni}, {Pineau}, {Plachy}, {Plum}, {Poujoulet},
  {Pr{\v{s}}a}, {Pulone}, {Ragaini}, {Rago}, {Rambaux}, {Ramos-Lerate},
  {Ranalli}, {Rauw}, {Read}, {Regibo}, {Renk}, {Reyl{\'e}}, {Ribeiro},
  {Rimoldini}, {Ripepi}, {Riva}, {Rixon}, {Roelens}, {Romero-G{\'o}mez},
  {Rowell}, {Royer}, {Rudolph}, {Ruiz-Dern}, {Sadowski}, {Sagrist{\`a}
  Sell{\'e}s}, {Sahlmann}, {Salgado}, {Salguero}, {Sarasso}, {Savietto},
  {Schnorhk}, {Schultheis}, {Sciacca}, {Segol}, {Segovia}, {Segransan},
  {Serpell}, {Shih}, {Smareglia}, {Smart}, {Smith}, {Solano}, {Solitro},
  {Sordo}, {Soria Nieto}, {Souchay}, {Spagna}, {Spoto}, {Stampa}, {Steele},
  {Steidelm{\"u}ller}, {Stephenson}, {Stoev}, {Suess}, {S{\"u}veges}, {Surdej},
  {Szabados}, {Szegedi-Elek}, {Tapiador}, {Taris}, {Tauran}, {Taylor},
  {Teixeira}, {Terrett}, {Tingley}, {Trager}, {Turon}, {Ulla}, {Utrilla},
  {Valentini}, {van Elteren}, {Van Hemelryck}, {van Leeuwen}, {Varadi},
  {Vecchiato}, {Veljanoski}, {Via}, {Vicente}, {Vogt}, {Voss}, {Votruba},
  {Voutsinas}, {Walmsley}, {Weiler}, {Weingrill}, {Werner}, {Wevers},
  {Whitehead}, {Wyrzykowski}, {Yoldas}, {{\v{Z}}erjal}, {Zucker}, {Zurbach},
  {Zwitter}, {Alecu}, {Allen}, {Allende Prieto}, {Amorim},
  {Anglada-Escud{\'e}}, {Arsenijevic}, {Azaz}, {Balm}, {Beck}, {Bernstein},
  {Bigot}, {Bijaoui}, {Blasco}, {Bonfigli}, {Bono}, {Boudreault}, {Bressan},
  {Brown}, {Brunet}, {Bunclark}, {Buonanno}, {Butkevich}, {Carret}, {Carrion},
  {Chemin}, {Ch{\'e}reau}, {Corcione}, {Darmigny}, {de Boer}, {de Teodoro}, {de
  Zeeuw}, {Delle Luche}, {Domingues}, {Dubath}, {Fodor}, {Fr{\'e}zouls},
  {Fries}, {Fustes}, {Fyfe}, {Gallardo}, {Gallegos}, {Gardiol}, {Gebran},
  {Gomboc}, {G{\'o}mez}, {Grux}, {Gueguen}, {Heyrovsky}, {Hoar}, {Iannicola},
  {Isasi Parache}, {Janotto}, {Joliet}, {Jonckheere}, {Keil}, {Kim},
  {Klagyivik}, {Klar}, {Knude}, {Kochukhov}, {Kolka}, {Kos}, {Kutka}, {Lainey},
  {LeBouquin}, {Liu}, {Loreggia}, {Makarov}, {Marseille}, {Martayan},
  {Martinez-Rubi}, {Massart}, {Meynadier}, {Mignot}, {Munari}, {Nguyen},
  {Nordlander}, {Ocvirk}, {O'Flaherty}, {Olias Sanz}, {Ortiz}, {Osorio},
  {Oszkiewicz}, {Ouzounis}, {Palmer}, {Park}, {Pasquato}, {Peltzer}, {Peralta},
  {P{\'e}turaud}, {Pieniluoma}, {Pigozzi}, {Poels}, {Prat}, {Prod'homme},
  {Raison}, {Rebordao}, {Risquez}, {Rocca-Volmerange}, {Rosen}, {Ruiz-Fuertes},
  {Russo}, {Sembay}, {Serraller Vizcaino}, {Short}, {Siebert}, {Silva},
  {Sinachopoulos}, {Slezak}, {Soffel}, {Sosnowska}, {Strai{\v{z}}ys}, {ter
  Linden}, {Terrell}, {Theil}, {Tiede}, {Troisi}, {Tsalmantza}, {Tur},
  {Vaccari}, {Vachier}, {Valles}, {Van Hamme}, {Veltz}, {Virtanen}, {Wallut},
  {Wichmann}, {Wilkinson}, {Ziaeepour}, \& {Zschocke}}]{prusti2016}
{Gaia Collaboration}, {Prusti}, T., {de Bruijne}, J.~H.~J., {et~al.} 2016,
  \aap, 595, A1, \dodoi{10.1051/0004-6361/201629272}

\bibitem[{{Gaia Collaboration} {et~al.}(2018{\natexlab{a}}){Gaia
  Collaboration}, {Babusiaux}, {van Leeuwen}, {Barstow}, {Jordi}, {Vallenari},
  {Bossini}, {Bressan}, {Cantat-Gaudin}, {van Leeuwen}, {Brown}, {Prusti}, {de
  Bruijne}, {Bailer-Jones}, {Biermann}, {Evans}, {Eyer}, {Jansen}, {Klioner},
  {Lammers}, {Lindegren}, {Luri}, {Mignard}, {Panem}, {Pourbaix}, {Randich},
  {Sartoretti}, {Siddiqui}, {Soubiran}, {Walton}, {Arenou}, {Bastian},
  {Cropper}, {Drimmel}, {Katz}, {Lattanzi}, {Bakker}, {Cacciari},
  {Casta{\~n}eda}, {Chaoul}, {Cheek}, {De Angeli}, {Fabricius}, {Guerra},
  {Holl}, {Masana}, {Messineo}, {Mowlavi}, {Nienartowicz}, {Panuzzo},
  {Portell}, {Riello}, {Seabroke}, {Tanga}, {Th{\'e}venin}, {Gracia-Abril},
  {Comoretto}, {Garcia-Reinaldos}, {Teyssier}, {Altmann}, {Andrae}, {Audard},
  {Bellas-Velidis}, {Benson}, {Berthier}, {Blomme}, {Burgess}, {Busso},
  {Carry}, {Cellino}, {Clementini}, {Clotet}, {Creevey}, {Davidson}, {De
  Ridder}, {Delchambre}, {Dell'Oro}, {Ducourant},
  {Fern{\'a}ndez-Hern{\'a}ndez}, {Fouesneau}, {Fr{\'e}mat}, {Galluccio},
  {Garc{\'\i}a-Torres}, {Gonz{\'a}lez-N{\'u}{\~n}ez}, {Gonz{\'a}lez-Vidal},
  {Gosset}, {Guy}, {Halbwachs}, {Hambly}, {Harrison}, {Hern{\'a}ndez},
  {Hestroffer}, {Hodgkin}, {Hutton}, {Jasniewicz}, {Jean-Antoine-Piccolo},
  {Jordan}, {Korn}, {Krone-Martins}, {Lanzafame}, {Lebzelter}, {L{\"o}ffler},
  {Manteiga}, {Marrese}, {Mart{\'\i}n-Fleitas}, {Moitinho}, {Mora}, {Muinonen},
  {Osinde}, {Pancino}, {Pauwels}, {Petit}, {Recio-Blanco}, {Richards},
  {Rimoldini}, {Robin}, {Sarro}, {Siopis}, {Smith}, {Sozzetti}, {S{\"u}veges},
  {Torra}, {van Reeven}, {Abbas}, {Abreu Aramburu}, {Accart}, {Aerts},
  {Altavilla}, {{\'A}lvarez}, {Alvarez}, {Alves}, {Anderson}, {Andrei},
  {Anglada Varela}, {Antiche}, {Antoja}, {Arcay}, {Astraatmadja}, {Bach},
  {Baker}, {Balaguer-N{\'u}{\~n}ez}, {Balm}, {Barache}, {Barata}, {Barbato},
  {Barblan}, {Barklem}, {Barrado}, {Barros}, {Bartholom{\'e} Mu{\~n}oz},
  {Bassilana}, {Becciani}, {Bellazzini}, {Berihuete}, {Bertone}, {Bianchi},
  {Bienaym{\'e}}, {Blanco-Cuaresma}, {Boch}, {Boeche}, {Bombrun}, {Borrachero},
  {Bouquillon}, {Bourda}, {Bragaglia}, {Bramante}, {Breddels}, {Brouillet},
  {Br{\"u}semeister}, {Brugaletta}, {Bucciarelli}, {Burlacu}, {Busonero},
  {Butkevich}, {Buzzi}, {Caffau}, {Cancelliere}, {Cannizzaro}, {Carballo},
  {Carlucci}, {Carrasco}, {Casamiquela}, {Castellani}, {Castro-Ginard},
  {Charlot}, {Chemin}, {Chiavassa}, {Cocozza}, {Costigan}, {Cowell}, {Crifo},
  {Crosta}, {Crowley}, {Cuypers}, {Dafonte}, {Damerdji}, {Dapergolas}, {David},
  {David}, {de Laverny}, {De Luise}, {De March}, {de Martino}, {de Souza}, {de
  Torres}, {Debosscher}, {del Pozo}, {Delbo}, {Delgado}, {Delgado}, {Diakite},
  {Diener}, {Distefano}, {Dolding}, {Drazinos}, {Dur{\'a}n}, {Edvardsson},
  {Enke}, {Eriksson}, {Esquej}, {Eynard Bontemps}, {Fabre}, {Fabrizio},
  {Faigler}, {Falc{\~a}o}, {Farr{\`a}s Casas}, {Federici}, {Fedorets},
  {Fernique}, {Figueras}, {Filippi}, {Findeisen}, {Fonti}, {Fraile}, {Fraser},
  {Fr{\'e}zouls}, {Gai}, {Galleti}, {Garabato}, {Garc{\'\i}a-Sedano},
  {Garofalo}, {Garralda}, {Gavel}, {Gavras}, {Gerssen}, {Geyer}, {Giacobbe},
  {Gilmore}, {Girona}, {Giuffrida}, {Glass}, {Gomes}, {Granvik}, {Gueguen},
  {Guerrier}, {Guiraud}, {Guti{\'e}}, {Haigron}, {Hatzidimitriou}, {Hauser},
  {Haywood}, {Heiter}, {Helmi}, {Heu}, {Hilger}, {Hobbs}, {Hofmann}, {Holland},
  {Huckle}, {Hypki}, {Icardi}, {Jan{\ss}en}, {Jevardat de Fombelle}, {Jonker},
  {Juh{\'a}sz}, {Julbe}, {Karampelas}, {Kewley}, {Klar}, {Kochoska}, {Kohley},
  {Kolenberg}, {Kontizas}, {Kontizas}, {Koposov}, {Kordopatis},
  {Kostrzewa-Rutkowska}, {Koubsky}, {Lambert}, {Lanza}, {Lasne}, {Lavigne}, {Le
  Fustec}, {Le Poncin-Lafitte}, {Lebreton}, {Leccia}, {Leclerc},
  {Lecoeur-Taibi}, {Lenhardt}, {Leroux}, {Liao}, {Licata}, {Lindstr{\o}m},
  {Lister}, {Livanou}, {Lobel}, {L{\'o}pez}, {Managau}, {Mann}, {Mantelet},
  {Marchal}, {Marchant}, {Marconi}, {Marinoni}, {Marschalk{\'o}}, {Marshall},
  {Martino}, {Marton}, {Mary}, {Massari}, {Matijevi{\v{c}}}, {Mazeh},
  {McMillan}, {Messina}, {Michalik}, {Millar}, {Molina}, {Molinaro},
  {Moln{\'a}r}, {Montegriffo}, {Mor}, {Morbidelli}, {Morel}, {Morris},
  {Mulone}, {Muraveva}, {Musella}, {Nelemans}, {Nicastro}, {Noval},
  {O'Mullane}, {Ord{\'e}novic}, {Ord{\'o}{\~n}ez-Blanco}, {Osborne}, {Pagani},
  {Pagano}, {Pailler}, {Palacin}, {Palaversa}, {Panahi}, {Pawlak},
  {Piersimoni}, {Pineau}, {Plachy}, {Plum}, {Poggio}, {Poujoulet},
  {Pr{\v{s}}a}, {Pulone}, {Racero}, {Ragaini}, {Rambaux}, {Ramos-Lerate},
  {Regibo}, {Reyl{\'e}}, {Riclet}, {Ripepi}, {Riva}, {Rivard}, {Rixon},
  {Roegiers}, {Roelens}, {Romero-G{\'o}mez}, {Rowell}, {Royer}, {Ruiz-Dern},
  {Sadowski}, {Sagrist{\`a} Sell{\'e}s}, {Sahlmann}, {Salgado}, {Salguero},
  {Sanna}, {Santana-Ros}, {Sarasso}, {Savietto}, {Schultheis}, {Sciacca},
  {Segol}, {Segovia}, {S{\'e}gransan}, {Shih}, {Siltala}, {Silva}, {Smart},
  {Smith}, {Solano}, {Solitro}, {Sordo}, {Soria Nieto}, {Souchay}, {Spagna},
  {Spoto}, {Stampa}, {Steele}, {Steidelm{\"u}ller}, {Stephenson}, {Stoev},
  {Suess}, {Surdej}, {Szabados}, {Szegedi-Elek}, {Tapiador}, {Taris}, {Tauran},
  {Taylor}, {Teixeira}, {Terrett}, {Teyssandier}, {Thuillot}, {Titarenko},
  {Torra Clotet}, {Turon}, {Ulla}, {Utrilla}, {Uzzi}, {Vaillant}, {Valentini},
  {Valette}, {van Elteren}, {Van Hemelryck}, {Vaschetto}, {Vecchiato},
  {Veljanoski}, {Viala}, {Vicente}, {Vogt}, {von Essen}, {Voss}, {Votruba},
  {Voutsinas}, {Walmsley}, {Weiler}, {Wertz}, {Wevers}, {Wyrzykowski},
  {Yoldas}, {{\v{Z}}erjal}, {Ziaeepour}, {Zorec}, {Zschocke}, {Zucker},
  {Zurbach}, \& {Zwitter}}]{babusiaux2018}
{Gaia Collaboration}, {Babusiaux}, C., {van Leeuwen}, F., {et~al.}
  2018{\natexlab{a}}, \aap, 616, A10, \dodoi{10.1051/0004-6361/201832843}

\bibitem[{{Gaia Collaboration} {et~al.}(2018{\natexlab{b}}){Gaia
  Collaboration}, {Brown}, {Vallenari}, {Prusti}, {de Bruijne}, {Babusiaux},
  {Bailer-Jones}, {Biermann}, {Evans}, {Eyer}, {Jansen}, {Jordi}, {Klioner},
  {Lammers}, {Lindegren}, {Luri}, {Mignard}, {Panem}, {Pourbaix}, {Randich},
  {Sartoretti}, {Siddiqui}, {Soubiran}, {van Leeuwen}, {Walton}, {Arenou},
  {Bastian}, {Cropper}, {Drimmel}, {Katz}, {Lattanzi}, {Bakker}, {Cacciari},
  {Casta{\~n}eda}, {Chaoul}, {Cheek}, {De Angeli}, {Fabricius}, {Guerra},
  {Holl}, {Masana}, {Messineo}, {Mowlavi}, {Nienartowicz}, {Panuzzo},
  {Portell}, {Riello}, {Seabroke}, {Tanga}, {Th{\'e}venin}, {Gracia-Abril},
  {Comoretto}, {Garcia-Reinaldos}, {Teyssier}, {Altmann}, {Andrae}, {Audard},
  {Bellas-Velidis}, {Benson}, {Berthier}, {Blomme}, {Burgess}, {Busso},
  {Carry}, {Cellino}, {Clementini}, {Clotet}, {Creevey}, {Davidson}, {De
  Ridder}, {Delchambre}, {Dell'Oro}, {Ducourant},
  {Fern{\'a}ndez-Hern{\'a}ndez}, {Fouesneau}, {Fr{\'e}mat}, {Galluccio},
  {Garc{\'\i}a-Torres}, {Gonz{\'a}lez-N{\'u}{\~n}ez}, {Gonz{\'a}lez-Vidal},
  {Gosset}, {Guy}, {Halbwachs}, {Hambly}, {Harrison}, {Hern{\'a}ndez},
  {Hestroffer}, {Hodgkin}, {Hutton}, {Jasniewicz}, {Jean-Antoine-Piccolo},
  {Jordan}, {Korn}, {Krone-Martins}, {Lanzafame}, {Lebzelter}, {L{\"o}ffler},
  {Manteiga}, {Marrese}, {Mart{\'\i}n-Fleitas}, {Moitinho}, {Mora}, {Muinonen},
  {Osinde}, {Pancino}, {Pauwels}, {Petit}, {Recio-Blanco}, {Richards},
  {Rimoldini}, {Robin}, {Sarro}, {Siopis}, {Smith}, {Sozzetti}, {S{\"u}veges},
  {Torra}, {van Reeven}, {Abbas}, {Abreu Aramburu}, {Accart}, {Aerts},
  {Altavilla}, {{\'A}lvarez}, {Alvarez}, {Alves}, {Anderson}, {Andrei},
  {Anglada Varela}, {Antiche}, {Antoja}, {Arcay}, {Astraatmadja}, {Bach},
  {Baker}, {Balaguer-N{\'u}{\~n}ez}, {Balm}, {Barache}, {Barata}, {Barbato},
  {Barblan}, {Barklem}, {Barrado}, {Barros}, {Barstow}, {Bartholom{\'e}
  Mu{\~n}oz}, {Bassilana}, {Becciani}, {Bellazzini}, {Berihuete}, {Bertone},
  {Bianchi}, {Bienaym{\'e}}, {Blanco-Cuaresma}, {Boch}, {Boeche}, {Bombrun},
  {Borrachero}, {Bossini}, {Bouquillon}, {Bourda}, {Bragaglia}, {Bramante},
  {Breddels}, {Bressan}, {Brouillet}, {Br{\"u}semeister}, {Brugaletta},
  {Bucciarelli}, {Burlacu}, {Busonero}, {Butkevich}, {Buzzi}, {Caffau},
  {Cancelliere}, {Cannizzaro}, {Cantat-Gaudin}, {Carballo}, {Carlucci},
  {Carrasco}, {Casamiquela}, {Castellani}, {Castro-Ginard}, {Charlot},
  {Chemin}, {Chiavassa}, {Cocozza}, {Costigan}, {Cowell}, {Crifo}, {Crosta},
  {Crowley}, {Cuypers}, {Dafonte}, {Damerdji}, {Dapergolas}, {David}, {David},
  {de Laverny}, {De Luise}, {De March}, {de Martino}, {de Souza}, {de Torres},
  {Debosscher}, {del Pozo}, {Delbo}, {Delgado}, {Delgado}, {Di Matteo},
  {Diakite}, {Diener}, {Distefano}, {Dolding}, {Drazinos}, {Dur{\'a}n},
  {Edvardsson}, {Enke}, {Eriksson}, {Esquej}, {Eynard Bontemps}, {Fabre},
  {Fabrizio}, {Faigler}, {Falc{\~a}o}, {Farr{\`a}s Casas}, {Federici},
  {Fedorets}, {Fernique}, {Figueras}, {Filippi}, {Findeisen}, {Fonti},
  {Fraile}, {Fraser}, {Fr{\'e}zouls}, {Gai}, {Galleti}, {Garabato},
  {Garc{\'\i}a-Sedano}, {Garofalo}, {Garralda}, {Gavel}, {Gavras}, {Gerssen},
  {Geyer}, {Giacobbe}, {Gilmore}, {Girona}, {Giuffrida}, {Glass}, {Gomes},
  {Granvik}, {Gueguen}, {Guerrier}, {Guiraud}, {Guti{\'e}rrez-S{\'a}nchez},
  {Haigron}, {Hatzidimitriou}, {Hauser}, {Haywood}, {Heiter}, {Helmi}, {Heu},
  {Hilger}, {Hobbs}, {Hofmann}, {Holland}, {Huckle}, {Hypki}, {Icardi},
  {Jan{\ss}en}, {Jevardat de Fombelle}, {Jonker}, {Juh{\'a}sz}, {Julbe},
  {Karampelas}, {Kewley}, {Klar}, {Kochoska}, {Kohley}, {Kolenberg},
  {Kontizas}, {Kontizas}, {Koposov}, {Kordopatis}, {Kostrzewa-Rutkowska},
  {Koubsky}, {Lambert}, {Lanza}, {Lasne}, {Lavigne}, {Le Fustec}, {Le
  Poncin-Lafitte}, {Lebreton}, {Leccia}, {Leclerc}, {Lecoeur-Taibi},
  {Lenhardt}, {Leroux}, {Liao}, {Licata}, {Lindstr{\o}m}, {Lister}, {Livanou},
  {Lobel}, {L{\'o}pez}, {Managau}, {Mann}, {Mantelet}, {Marchal}, {Marchant},
  {Marconi}, {Marinoni}, {Marschalk{\'o}}, {Marshall}, {Martino}, {Marton},
  {Mary}, {Massari}, {Matijevi{\v{c}}}, {Mazeh}, {McMillan}, {Messina},
  {Michalik}, {Millar}, {Molina}, {Molinaro}, {Moln{\'a}r}, {Montegriffo},
  {Mor}, {Morbidelli}, {Morel}, {Morris}, {Mulone}, {Muraveva}, {Musella},
  {Nelemans}, {Nicastro}, {Noval}, {O'Mullane}, {Ord{\'e}novic},
  {Ord{\'o}{\~n}ez-Blanco}, {Osborne}, {Pagani}, {Pagano}, {Pailler},
  {Palacin}, {Palaversa}, {Panahi}, {Pawlak}, {Piersimoni}, {Pineau}, {Plachy},
  {Plum}, {Poggio}, {Poujoulet}, {Pr{\v{s}}a}, {Pulone}, {Racero}, {Ragaini},
  {Rambaux}, {Ramos-Lerate}, {Regibo}, {Reyl{\'e}}, {Riclet}, {Ripepi}, {Riva},
  {Rivard}, {Rixon}, {Roegiers}, {Roelens}, {Romero-G{\'o}mez}, {Rowell},
  {Royer}, {Ruiz-Dern}, {Sadowski}, {Sagrist{\`a} Sell{\'e}s}, {Sahlmann},
  {Salgado}, {Salguero}, {Sanna}, {Santana-Ros}, {Sarasso}, {Savietto},
  {Schultheis}, {Sciacca}, {Segol}, {Segovia}, {S{\'e}gransan}, {Shih},
  {Siltala}, {Silva}, {Smart}, {Smith}, {Solano}, {Solitro}, {Sordo}, {Soria
  Nieto}, {Souchay}, {Spagna}, {Spoto}, {Stampa}, {Steele},
  {Steidelm{\"u}ller}, {Stephenson}, {Stoev}, {Suess}, {Surdej}, {Szabados},
  {Szegedi-Elek}, {Tapiador}, {Taris}, {Tauran}, {Taylor}, {Teixeira},
  {Terrett}, {Teyssandier}, {Thuillot}, {Titarenko}, {Torra Clotet}, {Turon},
  {Ulla}, {Utrilla}, {Uzzi}, {Vaillant}, {Valentini}, {Valette}, {van Elteren},
  {Van Hemelryck}, {van Leeuwen}, {Vaschetto}, {Vecchiato}, {Veljanoski},
  {Viala}, {Vicente}, {Vogt}, {von Essen}, {Voss}, {Votruba}, {Voutsinas},
  {Walmsley}, {Weiler}, {Wertz}, {Wevers}, {Wyrzykowski}, {Yoldas},
  {{\v{Z}}erjal}, {Ziaeepour}, {Zorec}, {Zschocke}, {Zucker}, {Zurbach}, \&
  {Zwitter}}]{brown2018}
{Gaia Collaboration}, {Brown}, A.~G.~A., {Vallenari}, A., {et~al.}
  2018{\natexlab{b}}, \aap, 616, A1, \dodoi{10.1051/0004-6361/201833051}

\bibitem[{{Gaia Collaboration} {et~al.}(2021){Gaia Collaboration}, {Brown},
  {Vallenari}, {Prusti}, {de Bruijne}, {Babusiaux}, {Biermann}, {Creevey},
  {Evans}, {Eyer}, {Hutton}, {Jansen}, {Jordi}, {Klioner}, {Lammers},
  {Lindegren}, {Luri}, {Mignard}, {Panem}, {Pourbaix}, {Randich}, {Sartoretti},
  {Soubiran}, {Walton}, {Arenou}, {Bailer-Jones}, {Bastian}, {Cropper},
  {Drimmel}, {Katz}, {Lattanzi}, {van Leeuwen}, {Bakker}, {Cacciari},
  {Casta{\~n}eda}, {De Angeli}, {Ducourant}, {Fabricius}, {Fouesneau},
  {Fr{\'e}mat}, {Guerra}, {Guerrier}, {Guiraud}, {Jean-Antoine Piccolo},
  {Masana}, {Messineo}, {Mowlavi}, {Nicolas}, {Nienartowicz}, {Pailler},
  {Panuzzo}, {Riclet}, {Roux}, {Seabroke}, {Sordo}, {Tanga}, {Th{\'e}venin},
  {Gracia-Abril}, {Portell}, {Teyssier}, {Altmann}, {Andrae}, {Bellas-Velidis},
  {Benson}, {Berthier}, {Blomme}, {Brugaletta}, {Burgess}, {Busso}, {Carry},
  {Cellino}, {Cheek}, {Clementini}, {Damerdji}, {Davidson}, {Delchambre},
  {Dell'Oro}, {Fern{\'a}ndez-Hern{\'a}ndez}, {Galluccio}, {Garc{\'\i}a-Lario},
  {Garcia-Reinaldos}, {Gonz{\'a}lez-N{\'u}{\~n}ez}, {Gosset}, {Haigron},
  {Halbwachs}, {Hambly}, {Harrison}, {Hatzidimitriou}, {Heiter},
  {Hern{\'a}ndez}, {Hestroffer}, {Hodgkin}, {Holl}, {Jan{\ss}en}, {Jevardat de
  Fombelle}, {Jordan}, {Krone-Martins}, {Lanzafame}, {L{\"o}ffler}, {Lorca},
  {Manteiga}, {Marchal}, {Marrese}, {Moitinho}, {Mora}, {Muinonen}, {Osborne},
  {Pancino}, {Pauwels}, {Petit}, {Recio-Blanco}, {Richards}, {Riello},
  {Rimoldini}, {Robin}, {Roegiers}, {Rybizki}, {Sarro}, {Siopis}, {Smith},
  {Sozzetti}, {Ulla}, {Utrilla}, {van Leeuwen}, {van Reeven}, {Abbas}, {Abreu
  Aramburu}, {Accart}, {Aerts}, {Aguado}, {Ajaj}, {Altavilla}, {{\'A}lvarez},
  {{\'A}lvarez Cid-Fuentes}, {Alves}, {Anderson}, {Anglada Varela}, {Antoja},
  {Audard}, {Baines}, {Baker}, {Balaguer-N{\'u}{\~n}ez}, {Balbinot}, {Balog},
  {Barache}, {Barbato}, {Barros}, {Barstow}, {Bartolom{\'e}}, {Bassilana},
  {Bauchet}, {Baudesson-Stella}, {Becciani}, {Bellazzini}, {Bernet}, {Bertone},
  {Bianchi}, {Blanco-Cuaresma}, {Boch}, {Bombrun}, {Bossini}, {Bouquillon},
  {Bragaglia}, {Bramante}, {Breedt}, {Bressan}, {Brouillet}, {Bucciarelli},
  {Burlacu}, {Busonero}, {Butkevich}, {Buzzi}, {Caffau}, {Cancelliere},
  {C{\'a}novas}, {Cantat-Gaudin}, {Carballo}, {Carlucci}, {Carnerero},
  {Carrasco}, {Casamiquela}, {Castellani}, {Castro-Ginard}, {Castro Sampol},
  {Chaoul}, {Charlot}, {Chemin}, {Chiavassa}, {Cioni}, {Comoretto}, {Cooper},
  {Cornez}, {Cowell}, {Crifo}, {Crosta}, {Crowley}, {Dafonte}, {Dapergolas},
  {David}, {David}, {de Laverny}, {De Luise}, {De March}, {De Ridder}, {de
  Souza}, {de Teodoro}, {de Torres}, {del Peloso}, {del Pozo}, {Delbo},
  {Delgado}, {Delgado}, {Delisle}, {Di Matteo}, {Diakite}, {Diener},
  {Distefano}, {Dolding}, {Eappachen}, {Edvardsson}, {Enke}, {Esquej}, {Fabre},
  {Fabrizio}, {Faigler}, {Fedorets}, {Fernique}, {Fienga}, {Figueras},
  {Fouron}, {Fragkoudi}, {Fraile}, {Franke}, {Gai}, {Garabato},
  {Garcia-Gutierrez}, {Garc{\'\i}a-Torres}, {Garofalo}, {Gavras}, {Gerlach},
  {Geyer}, {Giacobbe}, {Gilmore}, {Girona}, {Giuffrida}, {Gomel}, {Gomez},
  {Gonzalez-Santamaria}, {Gonz{\'a}lez-Vidal}, {Granvik},
  {Guti{\'e}rrez-S{\'a}nchez}, {Guy}, {Hauser}, {Haywood}, {Helmi}, {Hidalgo},
  {Hilger}, {H{\l}adczuk}, {Hobbs}, {Holland}, {Huckle}, {Jasniewicz},
  {Jonker}, {Juaristi Campillo}, {Julbe}, {Karbevska}, {Kervella}, {Khanna},
  {Kochoska}, {Kontizas}, {Kordopatis}, {Korn}, {Kostrzewa-Rutkowska},
  {Kruszy{\'n}ska}, {Lambert}, {Lanza}, {Lasne}, {Le Campion}, {Le Fustec},
  {Lebreton}, {Lebzelter}, {Leccia}, {Leclerc}, {Lecoeur-Taibi}, {Liao},
  {Licata}, {Lindstr{\o}m}, {Lister}, {Livanou}, {Lobel}, {Madrero Pardo},
  {Managau}, {Mann}, {Marchant}, {Marconi}, {Marcos Santos}, {Marinoni},
  {Marocco}, {Marshall}, {Martin Polo}, {Mart{\'\i}n-Fleitas}, {Masip},
  {Massari}, {Mastrobuono-Battisti}, {Mazeh}, {McMillan}, {Messina},
  {Michalik}, {Millar}, {Mints}, {Molina}, {Molinaro}, {Moln{\'a}r},
  {Montegriffo}, {Mor}, {Morbidelli}, {Morel}, {Morris}, {Mulone}, {Munoz},
  {Muraveva}, {Murphy}, {Musella}, {Noval}, {Ord{\'e}novic}, {Orr{\`u}},
  {Osinde}, {Pagani}, {Pagano}, {Palaversa}, {Palicio}, {Panahi}, {Pawlak},
  {Pe{\~n}alosa Esteller}, {Penttil{\"a}}, {Piersimoni}, {Pineau}, {Plachy},
  {Plum}, {Poggio}, {Poretti}, {Poujoulet}, {Pr{\v{s}}a}, {Pulone}, {Racero},
  {Ragaini}, {Rainer}, {Raiteri}, {Rambaux}, {Ramos}, {Ramos-Lerate}, {Re
  Fiorentin}, {Regibo}, {Reyl{\'e}}, {Ripepi}, {Riva}, {Rixon}, {Robichon},
  {Robin}, {Roelens}, {Rohrbasser}, {Romero-G{\'o}mez}, {Rowell}, {Royer},
  {Rybicki}, {Sadowski}, {Sagrist{\`a} Sell{\'e}s}, {Sahlmann}, {Salgado},
  {Salguero}, {Samaras}, {Sanchez Gimenez}, {Sanna}, {Santove{\~n}a},
  {Sarasso}, {Schultheis}, {Sciacca}, {Segol}, {Segovia}, {S{\'e}gransan},
  {Semeux}, {Shahaf}, {Siddiqui}, {Siebert}, {Siltala}, {Slezak}, {Smart},
  {Solano}, {Solitro}, {Souami}, {Souchay}, {Spagna}, {Spoto}, {Steele},
  {Steidelm{\"u}ller}, {Stephenson}, {S{\"u}veges}, {Szabados}, {Szegedi-Elek},
  {Taris}, {Tauran}, {Taylor}, {Teixeira}, {Thuillot}, {Tonello}, {Torra},
  {Torra}, {Turon}, {Unger}, {Vaillant}, {van Dillen}, {Vanel}, {Vecchiato},
  {Viala}, {Vicente}, {Voutsinas}, {Weiler}, {Wevers}, {Wyrzykowski}, {Yoldas},
  {Yvard}, {Zhao}, {Zorec}, {Zucker}, {Zurbach}, \& {Zwitter}}]{brown2021}
---. 2021, \aap, 649, A1, \dodoi{10.1051/0004-6361/202039657}

\bibitem[{{Gao}(2019)}]{gao2019}
{Gao}, X.-h. 2019, \mnras, 486, 5405, \dodoi{10.1093/mnras/stz1213}

\bibitem[{{Gossage} {et~al.}(2018){Gossage}, {Conroy}, {Dotter}, {Choi},
  {Rosenfield}, {Cargile}, \& {Dolphin}}]{gossage2018}
{Gossage}, S., {Conroy}, C., {Dotter}, A., {et~al.} 2018, \apj, 863, 67,
  \dodoi{10.3847/1538-4357/aad0a0}

\bibitem[{{Gunn} {et~al.}(1988){Gunn}, {Griffin}, {Griffin}, \&
  {Zimmerman}}]{gunn1988}
{Gunn}, J.~E., {Griffin}, R.~F., {Griffin}, R.~E.~M., \& {Zimmerman}, B.~A.
  1988, \aj, 96, 198, \dodoi{10.1086/114801}

\bibitem[{{Hanson}(1975)}]{hanson1975}
{Hanson}, R.~B. 1975, \aj, 80, 379, \dodoi{10.1086/111753}

\bibitem[{{Hao} {et~al.}(2022){Hao}, {Xu}, {Wu}, {Lin}, {Liu}, \&
  {Li}}]{hao2022}
{Hao}, C.~J., {Xu}, Y., {Wu}, Z.~Y., {et~al.} 2022, \aap, 660, A4,
  \dodoi{10.1051/0004-6361/202243091}

\bibitem[{{Hao} {et~al.}(2021){Hao}, {Xu}, {Hou}, {Bian}, {Li}, {Wu}, {He},
  {Li}, \& {Liu}}]{hao2021}
{Hao}, C.~J., {Xu}, Y., {Hou}, L.~G., {et~al.} 2021, \aap, 652, A102,
  \dodoi{10.1051/0004-6361/202140608}

\bibitem[{{Healy} {et~al.}(2021){Healy}, {McCullough}, \&
  {Schlaufman}}]{Healy2021}
{Healy}, B.~F., {McCullough}, P.~R., \& {Schlaufman}, K.~C. 2021, \apj, 923,
  23, \dodoi{10.3847/1538-4357/ac281d}

\bibitem[{{Holland} {et~al.}(2000){Holland}, {Jameson}, {Hodgkin}, {Davies}, \&
  {Pinfield}}]{holland2000}
{Holland}, K., {Jameson}, R.~F., {Hodgkin}, S., {Davies}, M.~B., \& {Pinfield},
  D. 2000, \mnras, 319, 956, \dodoi{10.1046/j.1365-8711.2000.03949.x}

\bibitem[{{Hou}(2021)}]{hou2021}
{Hou}, L.~G. 2021, \frass, 8, 103, \dodoi{10.3389/fspas.2021.671670}

\bibitem[{{Kamann} {et~al.}(2019){Kamann}, {Bastian}, {Gieles}, {Balbinot}, \&
  {H{\'e}nault-Brunet}}]{kamann2019}
{Kamann}, S., {Bastian}, N.~J., {Gieles}, M., {Balbinot}, E., \&
  {H{\'e}nault-Brunet}, V. 2019, \mnras, 483, 2197,
  \dodoi{10.1093/mnras/sty3144}

\bibitem[{{Khalaj} \& {Baumgardt}(2013)}]{khalaj2013}
{Khalaj}, P., \& {Baumgardt}, H. 2013, \mnras, 434, 3236,
  \dodoi{10.1093/mnras/stt1239}

\bibitem[{{Kharchenko} {et~al.}(2013){Kharchenko}, {Piskunov}, {Schilbach},
  {R{\"o}ser}, \& {Scholz}}]{kharchenko2013}
{Kharchenko}, N.~V., {Piskunov}, A.~E., {Schilbach}, E., {R{\"o}ser}, S., \&
  {Scholz}, R.~D. 2013, \aap, 558, A53, \dodoi{10.1051/0004-6361/201322302}

\bibitem[{{Klein Wassink}(1924)}]{kleinwassink1924}
{Klein Wassink}, W.~J. 1924, \bain, 2, 183

\bibitem[{{Klein Wassink}(1927)}]{kleinwassink1927}
---. 1927, PGro, 41, 1

\bibitem[{{Kraus} \& {Hillenbrand}(2007)}]{kraus2007}
{Kraus}, A.~L., \& {Hillenbrand}, L.~A. 2007, \aj, 134, 2340,
  \dodoi{10.1086/522831}

\bibitem[{{Kroupa} {et~al.}(2001){Kroupa}, {Aarseth}, \& {Hurley}}]{kroupa2001}
{Kroupa}, P., {Aarseth}, S., \& {Hurley}, J. 2001, \mnras, 321, 699,
  \dodoi{10.1046/j.1365-8711.2001.04050.x}

\bibitem[{{Lada} \& {Lada}(2003)}]{lada2003}
{Lada}, C.~J., \& {Lada}, E.~A. 2003, \araa, 41, 57,
  \dodoi{10.1146/annurev.astro.41.011802.094844}

\bibitem[{{Lanzoni} {et~al.}(2013){Lanzoni}, {Mucciarelli}, {Origlia},
  {Bellazzini}, {Ferraro}, {Valenti}, {Miocchi}, {Dalessandro}, {Pallanca}, \&
  {Massari}}]{lanzoni2013}
{Lanzoni}, B., {Mucciarelli}, A., {Origlia}, L., {et~al.} 2013, \apj, 769, 107,
  \dodoi{10.1088/0004-637X/769/2/107}

\bibitem[{{Lanzoni} {et~al.}(2018){Lanzoni}, {Ferraro}, {Mucciarelli},
  {Pallanca}, {Lapenna}, {Origlia}, {Dalessandro}, {Valenti}, {Bellazzini},
  {Tiongco}, {Varri}, {Vesperini}, \& {Beccari}}]{lanzoni2018}
{Lanzoni}, B., {Ferraro}, F.~R., {Mucciarelli}, A., {et~al.} 2018, \apj, 861,
  16, \dodoi{10.3847/1538-4357/aac26a}

\bibitem[{{Leanza} {et~al.}(2022){Leanza}, {Pallanca}, {Ferraro}, {Lanzoni},
  {Dalessandro}, {Origlia}, {Mucciarelli}, {Valenti}, {Tiongco}, {Varri}, \&
  {Vesperini}}]{leanza2022}
{Leanza}, S., {Pallanca}, C., {Ferraro}, F.~R., {et~al.} 2022, \apj, 929, 186,
  \dodoi{10.3847/1538-4357/ac5d4e}

\bibitem[{{Lodieu} {et~al.}(2019){Lodieu}, {P{\'e}rez-Garrido}, {Smart}, \&
  {Silvotti}}]{lodieu2019}
{Lodieu}, N., {P{\'e}rez-Garrido}, A., {Smart}, R.~L., \& {Silvotti}, R. 2019,
  \aap, 628, A66, \dodoi{10.1051/0004-6361/201935533}

\bibitem[{{Loktin} \& {Popov}(2020)}]{loktin2020}
{Loktin}, A.~V., \& {Popov}, A.~A. 2020, AN, 341, 638,
  \dodoi{10.1002/asna.202013687}

\bibitem[{{Mapelli}(2017)}]{mapelli2017}
{Mapelli}, M. 2017, \mnras, 467, 3255, \dodoi{10.1093/mnras/stx304}

\bibitem[{{Marino} {et~al.}(2018){Marino}, {Milone}, {Casagrande}, {Przybilla},
  {Balaguer-N{\'u}{\~n}ez}, {Di Criscienzo}, {Serenelli}, \&
  {Vilardell}}]{marino2018}
{Marino}, A.~F., {Milone}, A.~P., {Casagrande}, L., {et~al.} 2018, \apjl, 863,
  L33, \dodoi{10.3847/2041-8213/aad868}

\bibitem[{{Mermilliod}(1981)}]{mermilliod1981}
{Mermilliod}, J.~C. 1981, \aap, 97, 235

\bibitem[{{Mermilliod} {et~al.}(1990){Mermilliod}, {Weis}, {Duquennoy}, \&
  {Mayor}}]{mermilliod1990}
{Mermilliod}, J.~C., {Weis}, E.~W., {Duquennoy}, A., \& {Mayor}, M. 1990, \aap,
  235, 114

\bibitem[{{Perryman} {et~al.}(1998){Perryman}, {Brown}, {Lebreton}, {Gomez},
  {Turon}, {Cayrel de Strobel}, {Mermilliod}, {Robichon}, {Kovalevsky}, \&
  {Crifo}}]{perryman1998}
{Perryman}, M.~A.~C., {Brown}, A.~G.~A., {Lebreton}, Y., {et~al.} 1998, \aap,
  331, 81.
\newblock \doarXiv{astro-ph/9707253}

\bibitem[{{Pinsonneault} {et~al.}(1998){Pinsonneault}, {Stauffer}, {Soderblom},
  {King}, \& {Hanson}}]{pinsonneault1998}
{Pinsonneault}, M.~H., {Stauffer}, J., {Soderblom}, D.~R., {King}, J.~R., \&
  {Hanson}, R.~B. 1998, \apj, 504, 170, \dodoi{10.1086/306077}

\bibitem[{{Plummer}(1915)}]{plummer1915}
{Plummer}, H.~C. 1915, \mnras, 76, 107, \dodoi{10.1093/mnras/76.2.107}

\bibitem[{{Poggio} {et~al.}(2021){Poggio}, {Drimmel}, {Cantat-Gaudin}, {Ramos},
  {Ripepi}, {Zari}, {Andrae}, {Blomme}, {Chemin}, {Clementini}, {Figueras},
  {Fouesneau}, {Fr{\'e}mat}, {Lobel}, {Marshall}, {Muraveva}, \&
  {Romero-G{\'o}mez}}]{poggio2021}
{Poggio}, E., {Drimmel}, R., {Cantat-Gaudin}, T., {et~al.} 2021, \aap, 651,
  A104, \dodoi{10.1051/0004-6361/202140687}

\bibitem[{{Reid} {et~al.}(2019){Reid}, {Menten}, {Brunthaler}, {Zheng}, {Dame},
  {Xu}, {Li}, {Sakai}, {Wu}, {Immer}, {Zhang}, {Sanna}, {Moscadelli}, {Rygl},
  {Bartkiewicz}, {Hu}, {Quiroga-Nu{\~n}ez}, \& {van Langevelde}}]{reid2019}
{Reid}, M.~J., {Menten}, K.~M., {Brunthaler}, A., {et~al.} 2019, \apj, 885,
  131, \dodoi{10.3847/1538-4357/ab4a11}

\bibitem[{{R{\"o}ser} \& {Schilbach}(2019)}]{roser2019}
{R{\"o}ser}, S., \& {Schilbach}, E. 2019, \aap, 627, A4,
  \dodoi{10.1051/0004-6361/201935502}

\bibitem[{{Seabroke} {et~al.}(2021){Seabroke}, {Fabricius}, {Teyssier},
  {Sartoretti}, {Katz}, {Cropper}, {Antoja}, {Benson}, {Smith}, {Dolding},
  {Gosset}, {Panuzzo}, {Th{\'e}venin}, {Allende Prieto}, {Blomme}, {Guerrier},
  {Huckle}, {Jean-Antoine}, {Haigron}, {Marchal}, {Baker}, {Damerdji}, {David},
  {Fr{\'e}mat}, {Jan{\ss}en}, {Jasniewicz}, {Lobel}, {Samaras}, {Plum},
  {Soubiran}, {Vanel}, {Zwitter}, {Ajaj}, {Caffau}, {Chemin}, {Royer},
  {Brouillet}, {Crifo}, {Guy}, {Hambly}, {Leclerc}, {Mastrobuono-Battisti}, \&
  {Viala}}]{seabroke2021}
{Seabroke}, G.~M., {Fabricius}, C., {Teyssier}, D., {et~al.} 2021, \aap, 653,
  A160, \dodoi{10.1051/0004-6361/202141008}

\bibitem[{{Vandenberg} \& {Bridges}(1984)}]{vandenberg1984}
{Vandenberg}, D.~A., \& {Bridges}, T.~J. 1984, \apj, 278, 679,
  \dodoi{10.1086/161836}

\bibitem[{{Vereshchagin} \& Chupina(2013)}]{vereshchagin2013a}
{Vereshchagin}, S.~V., \& Chupina, N.~V. 2013, \an, 334, 892,
  \dodoi{10.1002/asna.201311938}

\bibitem[{{Vereshchagin} {et~al.}(2013){Vereshchagin}, {Reva}, \&
  {Chupina}}]{vereshchagin2013b}
{Vereshchagin}, S.~V., {Reva}, V.~G., \& {Chupina}, N.~V. 2013, \arep, 57, 52,
  \dodoi{10.1134/S1063772912120062}

\bibitem[{{Wang} {et~al.}(2014){Wang}, {Chen}, {Lin}, {Pandey}, {Huang},
  {Panwar}, {Lee}, {Tsai}, {Tang}, {Goldman}, {Burgett}, {Chambers}, {Draper},
  {Flewelling}, {Grav}, {Heasley}, {Hodapp}, {Huber}, {Jedicke}, {Kaiser},
  {Kudritzki}, {Luppino}, {Lupton}, {Magnier}, {Metcalfe}, {Monet}, {Morgan},
  {Onaka}, {Price}, {Stubbs}, {Sweeney}, {Tonry}, {Wainscoat}, \&
  {Waters}}]{wang2014}
{Wang}, P.~F., {Chen}, W.~P., {Lin}, C.~C., {et~al.} 2014, \apj, 784, 57,
  \dodoi{10.1088/0004-637X/784/1/57}

\bibitem[{{Wayman}(1967)}]{wayman1967}
{Wayman}, P.~A. 1967, \pasp, 79, 156, \dodoi{10.1086/128457}

\bibitem[{{Wu} {et~al.}(2009){Wu}, {Zhou}, {Ma}, \& {Du}}]{wu2009}
{Wu}, Z.-Y., {Zhou}, X., {Ma}, J., \& {Du}, C.-H. 2009, \mnras, 399, 2146,
  \dodoi{10.1111/j.1365-2966.2009.15416.x}

\bibitem[{{Xu} {et~al.}(2013){Xu}, {Li}, {Reid}, {Menten}, {Zheng},
  {Brunthaler}, {Moscadelli}, {Dame}, \& {Zhang}}]{xu2013}
{Xu}, Y., {Li}, J.~J., {Reid}, M.~J., {et~al.} 2013, \apj, 769, 15,
  \dodoi{10.1088/0004-637X/769/1/15}

\end{thebibliography}
\bibliographystyle{aasjournal}

\end{document}